\documentclass[letterpaper, 10 pt, conference]{IEEEtran}  

\IEEEoverridecommandlockouts                              

\usepackage{stfloats}
\usepackage{bbm}

\usepackage{graphicx} 
\usepackage{amsmath} 
\usepackage{amssymb}  

\usepackage{lipsum}
\usepackage{graphicx}
\usepackage[T1]{fontenc}
\usepackage{aecompl}
\usepackage{amsfonts}
\usepackage{bbm}
\usepackage{ifthen}
\usepackage{subcaption}
\usepackage{multirow}
\usepackage{ifthen}
\usepackage{cite}
\usepackage[usenames, dvipsnames]{color}
\usepackage{url}
\usepackage{hyperref}
\usepackage[hyphenbreaks]{breakurl}
\usepackage{mathtools}	
\usepackage[normalem]{ulem}
\usepackage[noend]{algorithm2e}
\RestyleAlgo{ruled}
\usepackage{algpseudocode}
\usepackage{tabularx}
\usepackage{hhline}

\usepackage{amsthm}
\usepackage{balance}

\newtheorem{theorem}{Theorem}
\newtheorem{lemma}{Lemma}

\newtheorem{defn}{Definition}

\newtheorem{remark}{Remark}
\newtheorem{assumption}{Assumption}

\usepackage{bbm}

\newcommand{\U}{\mathcal{U}}
\newcommand{\D}{\mathbb{R}^n}

\makeatletter
\newcommand*{\centerfloat}{%
  \parindent \z@
  \leftskip \z@ \@plus 1fil \@minus \textwidth
  \rightskip\leftskip
  \parfillskip \z@skip}
\makeatother

\usepackage{enumerate}
\usepackage[shortlabels]{enumitem}

\usepackage{ifthen}
\newboolean{showcomments}
\setboolean{showcomments}{true}
\usepackage[usenames,dvipsnames]{color}
\usepackage{todonotes}

\definecolor{bleudefrance}{rgb}{0.19, 0.55, 0.91}
\definecolor{ao(english)}{rgb}{0.0, 0.5, 0.0}

\newcommand{\addcite}[0]{\ifthenelse{\boolean{showcomments}}
{\textcolor{purple}{(add cite(s)) }}{}}%

\newcommand{\enrique}[1]{  \ifthenelse{\boolean{showcomments}}
{\todo[inline,color=bleudefrance, caption ={}]{Enrique: #1}}{}}
\newcommand{\emmargin}[1]{\ifthenelse{\boolean{showcomments}}{\marginpar{\color{bleudefrance}\tiny EM: #1}}{}}

\newcommand{\roy}[1]{  \ifthenelse{\boolean{showcomments}}
{\todo[inline,color=red, caption ={}]{Roy: #1}}{}}

\newboolean{showedits}
\setboolean{showedits}{true}
\usepackage[markup=underlined]{changes}
\definechangesauthor[color=bleudefrance]{EM}
\newcommand{\aem}[1]{
\ifthenelse{\boolean{showedits}}
{\added[id=EM]{#1}}
{\!#1\hspace{-4.75pt}}
}
\newcommand{\repem}[2]{
\ifthenelse{\boolean{showedits}}
{\replaced[id=EM]{#1}{#2}}
{\!#1\hspace{-4.75pt}}
}
\newcommand{\dem}[1]{
\ifthenelse{\boolean{showedits}}
{\deleted[id=EM]{#1}}
{}
}

\makeatletter
\if@todonotes@disabled

\else

\fi
\makeatother

\newboolean{arxiv}
\setboolean{arxiv}{true}
\newboolean{lemmas}
\setboolean{lemmas}{true}
\allowdisplaybreaks

\usepackage{setspace}
\newboolean{with-appendix}
\setboolean{with-appendix}{false}
\mathtoolsset{showonlyrefs=true}

\title{\huge Data-driven Practical Stabilization of Nonlinear Systems via Chain Policies: Sample Complexity and Incremental Learning}

\author{Roy Siegelmann and Enrique Mallada
\thanks{}
\thanks{
Roy Siegelmann is with the Department of Applied Mathematics and Statistics, Johns Hopkins Unijersity, 3400 N Charles St., Baltimore, MD 21218, USA. Email: \texttt{rsiege15@jhu.edu}. 
Enrique Mallada is with the Department of Electrical and Computer Engineering, Johns Hopkins University, 3400 N Charles St., Baltimore, MD 21218, USA. Email: \texttt{mallada@jhu.edu}. 
}
}

\begin{document}
\bstctlcite{MyBSTcontrol} 

\allowdisplaybreaks

\maketitle

\begin{abstract}
We propose a method for data-driven practical stabilization of nonlinear systems with provable guarantees, based on the concept of \emph{Nonparametric Chain Policies (NCPs)}. The approach employs a normalized nearest-neighbor rule to assign, at each state, a finite-duration control signal derived from stored data, after which the process repeats. 
Unlike recent works that model the system as linear, polynomial, or polynomial fraction, we only assume the system to be locally Lipschitz.
Our analysis build son the framework of Recurrent Lyapunov Functions (RLFs), which enable data-driven certification of (practical) stability using standard norm functions instead of requiring the explicit construction of a classical Lyapunov function. To extend this framework, we introduce the concept of Recurrent Control Lyapunov Functions (R-CLFs), which can certify the existence of an NCP that practically stabilizes an arbitrarily small $c$-neighborhood of an equilibrium point. 
We also provide an explicit sample complexity guarantee of $\mathcal{O}\!\left((3/\rho)^d \log(R/c)\right)$ number of trajectories—where $R$ is the domain radius, $d$ the state dimension, and $\rho$ a system-dependent constant. The proposed Chain Policies are nonparametric, thus allowing new verified data to be readily incorporated into the policy to either improve convergence rate or enlarge the certified region. Numerical experiments illustrate and validate these properties.
\end{abstract}

\medskip


\section{Introduction}\label{ct-intro}


\textit{Data-driven control methods} offer a novel paradigm for synthesizing controllers directly from trajectory observations, potentially bypassing the need for accurate system models while reducing  computational burden and conservativeness of classical control synthesis~\cite{zhou1996robust,sontag2013mathematical}. Recent years have witnessed significant progress in data-driven control. These approaches, as well as their level of maturity, depend considerably on the underlying system properties. For \emph{linear systems}, the field has substantially matured: LMI-based formulas~\cite{de2019formulas} and convex programs~\cite{coulson2019data,berberich2020data} can transform trajectories into stabilizing feedback controllers with robustness~\cite{de2019formulas,berberich2020data,berberich2020robust,van2020noisy}, performance~\cite{coulson2019data,berberich2020data,de2021low} and sample complexity guarantees~\cite{chen2021black,werner2024sample}. 

For \emph{nonlinear systems}, several approaches have been proposed, with methods highly dependent on the implicit assumptions made on the nonlinear system class and the control synthesis methodology. One prolific line of works considers dynamics formed from dictionary-based hypothesis classes—e.g., using polynomials~\cite{dai2020semi,guo2021data,el2025sum}, fractions of polynomials~\cite{strasser2021data}, or general nonlinear functions~\cite{monshizadeh2025versatile}—and formulate semidefinite programs that render policies with a wide variety of guarantees, including contraction-based stability~\cite{oliveira2025convex,hu2025enforcing,tsukamoto2021contraction} or robustness~\cite{el2025sum}. Other methods employ general learning techniques to learn models or policies and leverage intrinsic system properties to provide different guarantees, e.g., Koopman operator methods that exploit spectral properties~\cite{huang2020data}, sample complexity analysis for stochastic dynamics~\cite{chen2019sample}, and conformal prediction approaches for statistical robustness~\cite{hsu2025statistical}.

Despite the effectiveness of these methods in synthesizing controllers with guarantee, many questions remain unanswered. First, sample complexity guarantees are typically technique-dependent and do not provide clear understanding of how data requirements scale with explicit system properties, such as state dimension or attainable performance levels, or the specific hypothesis class considered. Second, computational complexity of optimization-based methods scales poorly with dictionary size and state dimension. Third, incorporating new data necessitates resolving the underlying optimization problem, often requiring complete recomputation and discarding previous work. As a result, there remains a need for flexible data-driven approaches that can adapt to new information without structural assumptions while providing transparent performance-data trade-offs.

To address these challenges, we introduce the concept of \emph{Nonparametric Chain Policies} (NCPs), a data-driven approach that requires only Lipschitz assumptions on the system dynamics while providing explicit sample complexity guarantees for practical stabilization. NCPs employ a normalized nearest-neighbor rule to assign finite-duration control signals from a stored library of verified trajectories, enabling direct use of data without parametric modeling or optimization re-solving when new data arrives. Our theoretical guarantees build on the framework of Recurrent Lyapunov Functions~\cite{sspm2023cdc,sspm2025tac}, which we extend here for the control setting by introducing here the notion of Recurrent Control Lyapunov Functions (RCLFs, Section \ref{ct-rclfs}).

\noindent
\textbf{Contributions.} Our approach offers three key advantages over existing methods:
\begin{enumerate}
\item \textbf{Explicit sample complexity:} NCPs achieve practical exponential stabilization using $\mathcal{O}\big((3/\rho)^d \log(R/\delta)\big)$ sample trajectories, with transparent scaling in dimension $d$, target radius $R$, precision $\delta$, and a system-dependent parameter $\rho$. 
\item \textbf{Incremental learning:} The nonparametric nature of NCPs allows for new verified data to be seamlessly incorporated to expand a certified region, or improve performance, without discarding previous guarantees or re-solving optimization problems.
\item \textbf{Performance-complexity trade-offs:} The framework explicitly controls the trade-off between sample requirements and performance through a user-specified parameter $\rho$
 that relates best achievable performance and the performance guaranteed by the NCP.
\end{enumerate}

\noindent
\textbf{Organization.} Section~\ref{ct-prelim} introduces preliminaries. Section~\ref{ct-rclfs} presents Recurrent Control Lyapunov Functions (R-CLFs) and stability guarantees. Section~\ref{ct-ncp} defines Nonparametric Chain Policies and establishes sample complexity results. Section~\ref{ct-numerical} demonstrates the approach on nonlinear benchmarks, and Section~\ref{ct-conclusions} concludes.

\noindent
\textbf{Notation.} $\|\cdot\|$ denotes an arbitrary norm on $\mathbb{R}^n$. Given $x \in \mathbb{R}^n$ and $r > 0$, we define the closed ball of radius $r$ centered at $x$ as
$B_r(x) := \{ y \in \mathbb{R}^n \mid \|y - x\| \leq r \}$. 
For a scalar $a \in \mathbb{R}$, we write $[a]_+ := \max\{a, 0\}$. For a set $S \subseteq \mathbb{R}^n$ and a point $x \in \mathbb{R}^n$, the signed distance from $x$ to $S$ is defined as
\[
\mathrm{sd}(x, S) := 
\begin{cases}
\inf_{y \in \partial S} \|y - x\|, & \text{if } x \notin S, \\
- \inf_{y \in \partial S} \|y - x\|, & \text{if } x \in S.
\end{cases}
\]
\section{Preliminaries}\label{ct-prelim}

We consider a nonlinear control system:
\begin{equation}\label{eq:control-system}
    \dot{x}(t) = f(x(t), u(t)),  
\end{equation}
with state \( x(t)\in \mathbb{R}^n \) and input \( u(t)\in U\subseteq\mathbb{R}^m \). We define
\[
    \mathcal{U}^{(a,b]} := \{ u : (a,b] \to U \mid u \text{ measurable} \},
\]
as the set of admissible control signals on interval $(a,b]$, and set \(\mathcal{U}:=\mathcal{U}^{(0,\infty)}\).  
Given \(u_0 \in \mathcal{U}^{(0,a]}\) and \(u_1 \in \mathcal{U}^{(0,b]}\), their concatenation \(u_0 u_1 \in \mathcal{U}^{(0,a+b]}\) is defined by
\[
(u_0 u_1)(t)=
\begin{cases}
    u_0(t), & t \in (0,a],\\[0pt]
    u_1(t), & t \in (a,a+b].
\end{cases}
\]
More generally, for a sequence of control signals $u_n\in\mathcal{U}^{(0,\tau_n]}$, with $\tau_n>0$, $\forall n\in \mathbb{N}$, we further use $u_{[n]}:=u_0u_1\dots u_{n}$, and $u_{[\infty]}=\lim_{n\to\infty}u_{[n]}$.
In some occasions we slightly abuse notation by using \(u\) interchangeably to represent instantaneous inputs in \(U\) and signals in \(\mathcal{U}^{(a,b]}\); the intended meaning will always be clear from context.

For an initial state \(x\in \mathbb{R}^n\) and control signal \(u\in \mathcal{U}^{(0,a]}\), we denote by \(\phi(t,x,u)\) the solution of~\eqref{eq:control-system} for $t\in(0,a]$. 
We further assume the following regularity conditions for~\eqref{eq:control-system}.
\begin{assumption}[Forward Completeness]\label{assump:forward-complete}
The solutions of the control system \eqref{eq:control-system} are \textbf{forward complete}. Specifically, for each initial condition \( x \in \mathbb{R}^n \) and every control signal \( u \in \mathcal{U} \), the trajectory \( \phi(t,x,u) \) exists and remains bounded for all \( t \geq 0 \).
\end{assumption}

\begin{assumption}[Uniform Lipschitz Continuity]\label{as:Lipschitz}
The vector field $f(x,u)$ of system~\eqref{eq:control-system} is locally Lipschitz continuous in $x$, uniformly with respect to $u$. More precisely, for every compact set $S \subseteq \mathbb{R}^n$, there exists a constant ${L}_S \geq 0$ such that
\[
\|f(y,u)-f(x,u)\|\leq {L}_S\|y-x\|,\quad\forall x,y\in S,\;\forall u\in U.
\]
\end{assumption}

\subsection{Practical Exponential Stabilizability}
In this work, we aspire to render an equilibrium point $x^*\in \D$ practical exponentially stable. 
\begin{defn}[Equilibrium Point]\label{defn:equilibrium}
A point \( x^* \in \mathbb{R}^n \) is an \textbf{equilibrium point} of system~\eqref{eq:control-system} if there exists a control input \( u^* \in U \) such that $f(x^*, u^*) = 0$. 
\end{defn}

\begin{defn}[(Practical) Exponential Stabilizability]\label{defn:exp-practical-stabilizability}
Let $S \subseteq \mathbb{R}^n$. The equilibrium $x^*$ of system~\eqref{eq:control-system} is said to be:
\begin{enumerate}[(i)]
    \item \textbf{Exponentially Stabilizable} on $S$ if for every $x \in S$, there exists a control signal $u \in \mathcal{U}$ satisfying
    \begin{equation}\label{eq:exp-stability}
        \|\phi(t,x,u) - x^*\| \leq K e^{-\lambda t} \|x - x^*\|, \quad \forall t \geq 0;
    \end{equation}

    \item \textbf{Practically Exponentially Stabilizable} on $S$ if for every $x \in S$, there exists a control signal $u \in \mathcal{U}$ satisfying
    \begin{equation}\label{eq:practical-exp-stability}
        \|\phi(t,x,u) - x^*\| \leq K e^{-\lambda t} \|x - x^*\| + c, \quad \forall t \geq 0,
    \end{equation}
\end{enumerate}
for constants $K\ge 1$, $\lambda > 0$, and $c \geq 0$.
\end{defn}

It is well known from the topological entropy literature that it is impossible to exponentially stabilize a system, i.e., achieve \eqref{eq:exp-stability},  using a finite number of control signals~\cite{colonius2012minimal}. We will therefore aim to enforce the weaker notion of practical exponential stability, i.e., \eqref{eq:practical-exp-stability}, which follows the terminology of \cite{hamzi2004practical,colonius2021entropy}.

\subsection{Recurrent Lyapunov Functions}

To provide guarantees for our data--driven stabilization framework, we build on
the theory of \emph{Recurrent Lyapunov Functions (RLFs)}~\cite{sspm2023cdc,sspm2025tac}.
Unlike classical Lyapunov functions, which require strict decrease along trajectories,
RLFs only require a decrease at a sequence of \emph{recurrent times}. This relaxation
broadens the class of certificates available, while still ensuring
exponential stability.

We begin with the notion of containment times, which we define for general
trajectories of the controlled system. 

\begin{defn}[Containment Times]\label{def:containment-times}
Given a set $S \subset \D$, an initial state $x \in \D$, and an input $u \in \U$,
the set of \emph{containment times} is
\[
T_S(x,u) := \{t \in \mathbb{R}_{>0} \mid \phi(t,x,u) \in S\}.
\]
For constants $a,b > 0$ we define
\[
T_S(x,u;a,b) := T_S(x,u) \cap (a,a+b],
\]
and for convenience $T_S(x,u;b) := T_S(x,u;0,b)$.
\end{defn}

We now recall the definition of an RLF in the \emph{autonomous} case,
where the trajectory is uniquely determined by the initial condition.
In this case we write $\phi(t,x)$ for the flow of the system.

\begin{defn}[Recurrent Lyapunov Function]\label{def:RLF}
Let $S \subset \D$ be a compact set with $x^* \in \mathrm{int}(S)$.
A continuous function $V:\D \to \mathbb{R}_{\geq 0}$ is called a
\emph{Recurrent Lyapunov Function (RLF)} over $S$ with rate $\alpha > 0$
and horizon $\tau > 0$ if 
\begin{equation}\label{eq:RLF}
    \min_{t \in T_S(x;\tau)} e^{\alpha t} V(\phi(t,x)) \;\leq\; V(x), \qquad\forall x\in S,
\end{equation}
where $T_S(x;\tau) := \{t \in (0,\tau] \mid \phi(t,x)\in S\}$.
\end{defn}


It will also be useful to characterize the set of points that can be reached within a finite interval of time.
\begin{defn} [Reachable Tube]\label{defn:reachable-set}  
For the control system \eqref{eq:control-system}, a constant $\tau > 0$, and a set $S \subset \D$, we denote the $\tau$-reachable tube from $S$ within $\tau$ units of time by 
$$\mathcal{R}^{\tau}(S) = \bigcup_{x \in S,u\in \U, t \in [0, \tau]} \{\phi(t,x,u)\}.$$
\end{defn}
\section{Recurrent Control Lyapunov Functions}\label{ct-rclfs}

As mentioned before,  our guarantees rely on the theory of Recurrent Lyapunov Functions (RLFs) from \cite{sspm2023cdc}. In this section, we extend this notion to the control setting, introducing  Recurrent Control Lyapunov Functions (R-CLFs), and illustrate how they can be used to certify practical stabilizability.
Though RLFs and R-CLFs have been shonw to to certify stability, asymptotic stability and exponential stability~\cite{sspm2023cdc,sspm2025tac}, our focus here is on practical exponential stability and thus we will use the following definition.

\begin{defn}[Recurrent Control Lyapunov Function (R-CLF)]\label{defn:RCLF}
Consider the control system~\eqref{eq:control-system} with equilibrium $x^*\in \mathbb{R}^n$. Let $S\subseteq \mathbb{R}^n$ be a set satisfying $x^*\in\mathrm{int}(S)$. A continuous function $V : \mathbb{R}^n \to \mathbb{R}_{\geq 0}$ is a \textbf{Recurrent Control Lyapunov Function (R-CLF)} over $S$ if the following conditions hold:
\begin{enumerate}[(i)]
    \item \textbf{Positive Definiteness and Linear Bounds}: There exist constants $a_1, a_2 > 0$ such that
    \begin{equation}\label{eq:RCLF-bounds}
        a_1\|x - x^*\| \leq V(x) \leq a_2\|x - x^*\|, \quad \forall x \in S.
    \end{equation}
    
    \item \textbf{Control $\alpha$-Exponential $(\tau,\delta)$-Recurrence}: There exist constants $\tau, \alpha > 0$ and $\delta \geq 0$ such that for every $x \in S$, there exists $u \in \mathcal{U}^{[0,\tau)}$ satisfying
    \begin{equation}\label{eq:RCLF-recurrence}
        \min_{t \in T_{S}(x,u; \tau)}\!\!\! \!\!e^{\alpha t}(V(\phi(t,x,u)) \!-\! \delta) \leq [V(x) \!-\! \delta]_+\,.
    \end{equation}

\end{enumerate}
\end{defn}

The following lemma characterizes the long term behavior of the control system \eqref{eq:control-system} under the controls $u\in\U$ that are build upon concatenation of controls satisfying property \textit{(ii)} of Definition \ref{defn:RCLF}.
\begin{lemma}[Characterization of R-CLF]\label{lem:RCLF-charac}
Let assumptions \ref{assump:forward-complete} and \ref{as:Lipschitz} hold. Consider an equilibrium $x^*$ of \eqref{eq:control-system}  and a \emph{compact} set $S$ satisfying $x^*\in \mathrm{int}(S)$.
A function $V:\mathbb{R}^n\to \mathbb{R}_{\geq0}$ satisfying \eqref{eq:RCLF-bounds} is a Recurrent Control Lyapunov Function (R-CLF) over $S$  \emph{if and only if} there exists parameters $\alpha,\tau>0$ and $\delta\geq0$ such that for any $x\in S$ there is a sequence 
$\{t_n\}_{n\in\mathbb{N}}$ and $u\in\mathcal{U}$ satisfying the following conditions:
\begin{subequations}\label{eq:sequence-condition}
\begin{align}
    &\lim_{n\rightarrow+\infty} t_n = +\infty, \text{ with }\;\; t_{n+1} - t_n \in (0,\tau], \label{eq:sequence-condition-1} \\
    &\;\;\phi(t_n,x,u)\in S \,,\;\;\label{eq:sequence-condition-2}
    \text{ and } \\ 
    &\;\;V(\phi(t_n,x,u))-\delta\leq \begin{cases}
    e^{-\alpha t_n}(V(x)-\delta)\,,& n\leq \bar n\,,\\
    0 \,,&\text{ o.w.\,,}
    \end{cases}\label{eq:sequence-condition-3}
\end{align}
for a non necessarily finite $\bar n\in \mathbb{N}\cup\{\infty\}$.
\end{subequations}
\end{lemma}

\ifthenelse{\boolean{arxiv}}{
\begin{proof}[Proof of Lemma~\ref{lem:RCLF-charac}]
We prove each direction separately.

\noindent\textbf{Necessity ($\Rightarrow$):} 
Suppose that $V$ is a Recurrent Control Lyapunov Function (R-CLF) over the compact set $S$. By Definition~\ref{defn:RCLF}, there exist constants $\alpha,\tau>0$ and $\delta\geq0$ such that, for any $x\in S$, there exists $\bar u\in\mathcal{U}^{[0,\tau)}$ satisfying~\eqref{eq:RCLF-recurrence}. 

We will build $u\in\mathcal{U}$ and the sequence $\{t_n\}_{n\in\mathbb{N}}$ inductively. Let $t_0=0$, $x_0:=x\in S$, and define  for $n\geq 0$, 
\begin{align}
&\tau_{n} :=  \max\left\{\underset{{t \in T_{S}(x_n,\bar u_n;\tau)}}{\arg\min} \!e^{\alpha t}(V(\phi(t,x_n,\bar u_n))\!-\!\delta)\right\}\,,\label{eq:tau_n}\\
&t_{n+1} := t_{n} + \tau_n\,,\quad\text{and}\quad 
x_{n+1}: = \phi(\tau_{n}, x_n, u_n)\,,
\end{align}
where $\bar u_n\in \mathcal{U}^{[0,\tau)}$ is a control satisfying \eqref{eq:RCLF-recurrence}, $u_n$ is its restriction to the interval $(0,\tau_n]$, and $x_n\in S$ $\forall n\in\mathbb{N}$, by  definition. 
Next, let $u:=\lim_{t\rightarrow\infty} u_{[n]}$. Note that $\phi(t_0,x,u)=x_0\in S$, and whenever for some $n$, $\phi(t_n,x,u)=x_n\in S$ that by the group property of the flow 
\begin{align}
    \phi(t_{n+1},x,u)&=\phi(t_{n+1}-t_n,\phi(t_n,x,u),u_n)\\
    &=\phi(\tau_n,x_n,u_n)=x_{n+1}\in S\,,\label{eq:flow-property}
\end{align}
which by induction ensures that \eqref{eq:sequence-condition-2} holds.

Next, from the recurrence condition~\eqref{eq:RCLF-recurrence}, it follows that
as long as $V(\phi(t_n,x,u))\geq\delta$, then
$e^{\alpha\tau_n} \left(V(\phi(\tau_{n},x_n,\bar u_n)) \!-\!\delta\right) \!\leq\!  V(x_n)\!-\!\delta$, 
which implies 
\begin{align}
\!\!\!\!\!e^{\alpha t_{n+1}}(V(\phi(t_{n+1},x,u))\!-\!\delta)  &\!\leq\! e^{\alpha t_{n}} (V(\phi(t_{n},x,u))\!-\!\delta)\,,\label{eq:one-step-condition}
\end{align}
and, in particular, 
\begin{align}
e^{\alpha t_{n}} \left(V(\phi(t_{n},x,u)) -\delta\right) \!\leq\!  V(x)-\delta,\quad \forall n\leq\bar n\,,\label{eq:exponntial decrease sequence}
\end{align}
where $\bar n$ is the last instance with $V(x_{\bar n})\geq \delta$.
It also follows from \eqref{eq:RCLF-recurrence} that, when $\bar n<\infty$, for all $n\geq\bar n+1$, $V(\phi(t_{n},x,u))\leq \delta$, which completes \eqref{eq:sequence-condition-3}.

To show \eqref{eq:sequence-condition-1},  we first note that by definition, $t_{n+1}-t_n=\tau_n \in (0,\tau]$. Next, we will show that $t^*=\infty$. Suppose not, i.e., $t^*<\infty$. 
By continuity of $\phi(t,x,u)$ and compactness of $S$, $\phi(t_n,x,u)\to \phi(t^*,x,u)\in S$. 

Now, let $v_n:=V(x_{n})-\delta$ and $v^*=V(\phi(t^*,x,u))-\delta$. If $v^*>0$,
it follows from \eqref{eq:one-step-condition} and the continuity of $V$ that $v_n\downarrow v^*:=V(\phi(t^*,x,u)-\delta)$, and 
$e^{\alpha t^*}v^* \leq e^{\alpha t_{n+1}}v_{n+1}\leq e^{\alpha t_n}v_{n}\,.$
Thus, for large enough $n$, $t^*\in (t_n,t_n+\tau]$ and $t^*>t_{n+1}$, which implies 
\[
e^{\alpha t^*-t_n}v^* \leq e^{\alpha \tau_{n}}v_{n+1}\leq v_{n}\,,
\]
which contradicts $\tau_n$ being the max in \eqref{eq:tau_n}.
As similar argument holds when $v^*<0$, and thus $t^*=\infty$.

\noindent\textbf{Sufficiency ($\Leftarrow$):} 
To show that $V$ is a R-CLF, it is sufficient for any $x\in S$ to restrict the corresponding $u\in \mathcal{U}$ that satisfies \eqref{eq:sequence-condition} to the interval $[0,\tau)$ and choosing $t_1$ from the sequence \eqref{eq:sequence-condition} to show \eqref{eq:RCLF-recurrence}.
\end{proof}
}{
\noindent \textit{Proof.}~The proof is omitted due to page limits. \hfill \qed
}

We will leverage Lemma \ref{lem:RCLF-charac} to prove (practical) exponential convergence of trajectories. To that end, we need to bound how much a trajectory can travel in between instance of exponential convergence \eqref{eq:sequence-condition-3}. The following lemma provides a mechanism to obtain such bounds. The proof is based on Grönwall's Lemma \cite[Lemma A.1]{Khalil2002} and can be found in \cite{sm2025nahs}.

\begin{lemma}[Containment Lemma]\label{lem:containment-control}
    Let Assumption \ref{as:Lipschitz} hold. Consider a compact set  $S \subset \D
    $ and a constant $\tau>0$. Then, for any $x \in S, u \in \mathcal{U}$ the following holds:
    \begin{equation}
        \max_{t\in[0,\tau]} \mathrm{d}(\phi(t,x,u),S) \leq F_S \,\tau e^{L \tau}
    \end{equation}
    where $ L :=L_{\mathcal{R}^{\tau}(S)}$.
\end{lemma}

We are now ready to show that R-CLFs as defined in Definition \ref{defn:RCLF} guarantee exponential stabilizability.
\begin{theorem}[R-CLF Implies (Practical) Exponential Stabilizability]\label{thm:RCLF-stabilizability}
Consider the control system~\eqref{eq:control-system} with equilibrium $x^*\in \mathbb{R}^n$, and let $S$ be a set satisfying $S\subseteq \mathbb{R}^n$ and $x^*\in\mathrm{int}(S)$. Let Assumption \ref{assump:forward-complete} and Assumption \ref{as:Lipschitz} hold, and $V: \mathbb{R}^n \to \mathbb{R}_{\geq 0}$ be a Recurrent Control Lyapunov Function over $S$, with constants $\alpha,\tau>0$, $\delta\geq0$, and linear bound constants $a_1,a_2>0$ from~\eqref{eq:RCLF-bounds}. 

Then, the equilibrium $x^*$ is (practically) exponentially stabilizable on $S$ (when $\delta>0$). In particular, for every initial condition $x \in S$, a control signal $u\in\mathcal{U}$ satisfying \eqref{eq:sequence-condition} for some sequence of times $\{t_n\}_{n\in\mathbb{N}}$ ensures:
\begin{equation}\label{eq:RCLF-stabilizability}
    \|\phi(t,x,u)-x^*\|\leq K e^{-\lambda t}\|x-x^*\| + c,\quad \forall t\geq0,
\end{equation}
where
\begin{equation}
    \lambda \!:=\! \alpha, \; K\!:=\!\frac{a_2}{a_1} e^{\alpha\tau}(1\!+\!L\tau e^{L\tau}), \; \text{ and } \; \textcolor{black}{c\!:=\! \frac{\delta}{a_1} (1\!+\!L\tau e^{L\tau})},
\end{equation}
with \(L\!\!:=\!\!L_{R^{\tau}(S)}\). 
\end{theorem}

\ifthenelse{\boolean{arxiv}}{
\begin{proof}
We will use the control $u\in\mathcal{U}$ from Lemma~\ref{lem:RCLF-charac} to prove this theorem.
Given \(x \in S\), by Lemma \ref{lem:RCLF-charac}, there exists $u\in\U$ and a sequence $t_n, x_n:=\phi(t_n,x)\in S$ satisfying \eqref{eq:sequence-condition}. There are two cases. First, assume that $n \leq \bar{n}\in \mathbb{N}\cup \{\infty\}$. Since \(V(x_n) \leq e^{-\alpha t_n}(V(x) - \delta) + \delta\) and \(V(x_n) \geq a_1 \|x_n - x^*\|\), it follows that
\[
r_n:=\|x_n - x^*\| \leq \frac{1}{a_1} V(x_n)
\leq \frac{a_2}{a_1} e^{-\alpha t_n} \|x - x^*\| + \frac{\delta}{a_1}.
\]
Consider any time \(t \in (t_n, t_{n+1}]\) and $B_n:=B_n(x^*)\cap S$. By applying the containment lemma on $B_n$ and using the fact that $L\geq L_{\mathcal{R}^\tau(B_n)}$, we get, using the triangle inequality:
\begin{align*}
\|\phi(t,x,u) - x^*\| &\leq \|x_n - x^*\|+\|\phi(t,x,u) - x_n\| \\
    &\leq r_n + F_{r_n}\tau e^{L\tau}\leq (1+L\tau e^{L\tau})r_n
\end{align*}
where $r_n:=||x_n-x^*||$.
Then for any $n\leq \bar n$ and any $t\in(t_n,t_{n+1}]$ we have,
\begin{align}
    & \|\phi(t,x,u) - x^*\| \leq 
   (1+L\tau e^{L\tau})r_n \\
   &\leq  (1+L\tau e^{L\tau})\left(\frac{a_2}{a_1} e^{-\alpha t_n} \|x - x^*\| + \frac{\delta}{a_1}\right)\\
    & \leq Ke^{-\alpha \tau}e^{-\alpha t_n}
    \|x - x^*\| + \frac{\delta}{a_1}(1+L\tau e^{L\tau})\\
    &\leq Ke^{-\alpha t}||x-x^*|| + c
\end{align}
where the last step follows, since \(t \leq t_{n+1} \leq t_n + \tau\), which implies  \(-t_n -\tau \leq - t\), so that \(
 e^{-\alpha \tau} e^{-\alpha t_n} \leq  e^{-\alpha t}\). 

If $\bar n=\infty$ we are done.  Otherwise, consider \(n > \bar n\). We have
\(r_n=\|x_n - x^*\| \leq \frac{V(x_n)}{a_1} \leq \frac{\delta}{a_1}\) $\forall n>\bar n$, and thus by Lemma~\ref{lem:containment-control} again, for all $t > t_{\bar{n}}$
\[
\|\phi(t,x,u) - x^*\| \leq (1+L\tau e^{L\tau})r_n \leq (1+L\tau e^{L\tau})
\frac{\delta}{a_1}  = c.
\]
Thus, for all \(t \geq 0\),
\[
\|\phi(t,x,u) - x^*\| \leq K e^{-\lambda t} \|x - x^*\| + c,
\]
as desired.
\end{proof}
}{
\noindent \textit{Proof.}~The proof is omitted due to page limits. \hfill \qed
}

Theorem~\ref{thm:RCLF-stabilizability} states that the existence of an R-CLF implies that $x^*$ can be made practically exponentially stable. At the core of its proof is the fact that one can find a function $V$ that satisfies the recurrent condition \eqref{eq:RCLF-recurrence}. A key observation of \cite{sspm2025tac}, is that condition \eqref{eq:RCLF-recurrence} can be met by a norm, provided $\tau$ and $\alpha$ are properly chosen (c.f. \cite[Theorem 6]{sspm2025tac}). The the caveat is, however, that in order to make R-CLFs practically useful, on would need to store, for each $x\in S$, a suitable $u:[0,\tau)\to U$ that ensures \eqref{eq:RCLF-recurrence}. In the next section we surprisingly show that when $\delta>0$, only a finite number of such signals are needed.

\section{Non-Parametric Chain Policies}\label{ct-ncp}

In the previous section, we introduced Recurrent Control Lyapunov Functions (R-CLFs) to characterize exponential stabilizability via carefully selected control signals. In this section, we propose \emph{nonparametric chain policies}, a systematic, data-driven approach for generating these stabilizing signals. The proposed method aligns closely with recent developments in topological entropy—a notion quantifying the minimal complexity required to accomplish various control tasks (see, e.g., \cite{colonius2009invariance,colonius2012minimal,colonius2021entropy}). A distinctive feature of our method is that we do not assume the control signals can be generated online; instead, we explicitly store them in a finite set, called a \emph{control alphabet}~\cite{sm2025nahs}.

\begin{defn}[Control Alphabet]\label{defn:control-alphabet}
A \textbf{control alphabet} is a finite collection of control signals
\begin{equation}\label{eq:alphabet}
    \mathcal{A} := \{v_i:(0,\tau_i]\to U\}_{i=0}^M,
\end{equation}
where each \(v_i\) is piecewise continuous and defined over a duration \(\tau_i>0\).
\end{defn}

The control alphabet provides a library of candidate signals. To deploy them, we assign specific controls to regions of influence within the state space. To aid this task we define an assignment set.

\begin{defn}[Assignment Set]\label{defn:assignment-set}
An \textbf{assignment set} is a finite collection of verification triples
\[
\mathcal{K} := \{(x_i,r_i,v_i)\}_{i=1}^N \subseteq \mathbb{R}^n \times \mathbb{R}_{>0} \times \mathcal{A},
\]
where $x_i\in\mathbb{R}^n$ is a center point, $r_i>0$ is its radius, and $v_i\in\mathcal{A}$ is the control signal assigned to that region. 
The \emph{support} of $\mathcal{K}$ is
\[
\mathrm{Supp}(\mathcal{K}) := \bigcup_{i=1}^N B_{r_i}(x_i).
\]
We denote $N:=|\mathcal{K}|$ as the size of the assignment set. 
\end{defn}

While an assignment set specifies regions of influence, it does not by itself resolve 
which control to apply when balls overlap, nor what to do when a state lies outside 
$\mathrm{Supp}(\mathcal{K})$. To address this, we introduce a normalized 
nearest-neighbor rule with a fall-back option:
\begin{equation}\label{eq:ratio-K}
\iota_{\mathcal{K}}(x) :=
\begin{cases}
\arg\min_{i:(x_i,r_i,v_i)\in\mathcal{K}} \dfrac{\|x-x_i\|}{r_i}, & r_{\mathcal{K}}(x)\le 1,\\[2pt]
0, & \text{otherwise},
\end{cases}
\end{equation}
where 
\[
r_{\mathcal{K}}(x) := \min_{(x_i,r_i,v_i)\in\mathcal{K}} \frac{\|x-x_i\|}{r_i},
\]
and $\iota_{\mathcal{K}}(x)=0$ corresponds to selecting the default control $v_0$.

\begin{remark}
We designate \(v_0\in\mathcal{A}\) as the \emph{default control}. Unless otherwise stated, we take $v_0(t)=u^*\in U$, $\forall t\in[0,\tau_0)$, where $u^*$ is the equilibrium control of Definition~\ref{defn:equilibrium}.
\end{remark}

The index map $\iota_{\mathcal{K}}$ specifies, for any state $x$, which control from the assignment set (or the default control) should be applied. Building on this rule, we can now formalize the induced feedback policy.

\begin{defn}[Nonparametric Chain Policy]\label{defn:npc-policy}
Given an assignment set $\mathcal{K}$ and default control $v_0$, the 
\textbf{nonparametric chain policy} is given by the map $\pi_\mathcal{K}: \mathbb{R}^n\to\mathcal{A}$:
\begin{equation}\label{eq:chain-policy-K}
\pi_{\mathcal{K}}(x) := v_{\iota_{\mathcal{K}}(x)}.
\end{equation}
\end{defn}

\begin{remark}
The policy $\pi_{\mathcal{K}}$ induces an infinite-horizon control 
signal $u_{\mathcal{K},x}\in\mathcal{U}$ through concatenation. 
Starting with $x_0=x$ and the empty signal $u_{[0]}=\emptyset$, for each $n\ge 0$ define
\begin{equation}\label{eq:npc-sequence-K}
u_{[n+1]} = u_{[n]} \, v_{\iota_{\mathcal{K}}(x_n)},\;\;
x_{n+1} = \phi\big(\tau_{\iota_{\mathcal{K}}(x_n)},\,x_n,\,v_{\iota_{\mathcal{K}}(x_n)}\big).
\end{equation}
The resulting control is then
\[
u_{\mathcal{K},x} := \lim_{n\to\infty} u_{[n]}.
\]
\end{remark}

\subsection{Convergence Guarantees of NCPs}
With the nonparametric chain policy in place, we now turn to its stability properties. 
The following theorem establishes conditions under which such a policy renders the 
equilibrium $x^*$ practically exponentially stable on a prescribed region.

\begin{theorem}[Practical Exponential Stabilization via Chain Policies]\label{thm:NPC}
Consider an equilibrium point $x^* \in \mathbb{R}^n$ of \eqref{eq:control-system}, 
and let $S \subseteq \mathbb{R}^n$ be a set with $x^* \in \mathrm{int}(S)$. 
Let $\pi_{\mathcal{K}}$ denote a nonparametric chain policy associated with the 
assignment set $\mathcal{K} = \{(x_i,r_i,v_i)\}_{i=1}^N$ and a default control 
$v_0 \in \mathcal{A}$, and define $\tau := \max\{\tau_0,\tau_1,\dots,\tau_N\}$, and 
let $L := L_{\mathcal{R}^{\tau}(S)}$.
Suppose the following hold:
\begin{enumerate}[(i)]
    \item \textbf{Covering.} 
    There exists $\varepsilon>0$ such that
    \begin{subequations}\label{eq:covering}
        \begin{gather}
        B_\varepsilon(x^*) \subset B_{\varepsilon(1+L\tau e^{L\tau})}(x^*) \subset \mathrm{int}(S), \label{eq:covering-epsilon}\\
        \mathrm{cl}(S\setminus B_\varepsilon(x^*)) \subseteq \mathrm{Supp}(\mathcal{K}). \label{eq:covering-support}
        \end{gather}
    \end{subequations}
    
    \item \textbf{Verification.} 
    For each $(x_i,r_i,v_i)\in\mathcal{K}$ with $\tau_i>0$,
    \begin{subequations}\label{eq:verification-condition}
    \begin{align}
\!\!\!\!\!\!\!\!\!\!\!\!\!    e^{\alpha \tau_i}\bigl(\|\phi(\tau_i,x_i,v_i)\!-\!x^*\| \!+\! r_i e^{L\tau_i}\bigr)
        &\!\leq \|x_i \!-\! x^*\|\!-\!r_i, \label{eq:verify-exp-decreasing}\\[2pt]
    \mathrm{sd}(\phi(\tau_i,x_i,v_i),S) + r_i e^{L\tau_i} &\!\leq 0, \label{eq:verify-feasibility}
    \end{align}
    \end{subequations}
    \item \textbf{Equilibrium.} 
    For all $t\in[0,\tau_0)$, $\phi(t, x^*, v_0) = x^*$.
\end{enumerate}

Then the equilibrium $x^*$ is practically exponentially stable on $S$ under 
the policy $\pi_{\mathcal{K}}$, with constants
\[
\lambda=\alpha, \qquad 
K= e^{\alpha\tau}(1+L\tau e^{L\tau}), \qquad
c=\varepsilon(1+L\tau e^{L\tau}).
\]
\end{theorem}
\begin{proof}
We will show that the control $u_{\mathcal{K},x}$ induced by the nonparametric chain policy $\pi_{\mathcal{K}}$ admits a sequence of times $\{t_n\}_{n\in\mathbb{N}}$ that satisfies the conditions of Lemma~\ref{lem:RCLF-charac} for the function $V(x) = \|x - x^*\|$ over the set $S$. This establishes two points:  
(1) $V=\|\cdot-x^*\|$ is an R-CLF with rate $\alpha$, and  
(2) the control $u_{\mathcal{K},x}$ practically stabilizes $x^*$ with exponential rate $\alpha$ over $S$.

Let $x_0 = x\in S$, $t_0=0$, and $u_{[0]}=\emptyset$. Define the sequences $\{x_n\}$, $\{t_n\}$, and $u_{[n]}$ according to~\eqref{eq:npc-sequence-K}, i.e.,
\begin{align}
    &x_{n+1} = \phi\big(\tau_{\iota_{\mathcal{K}}(x_n)},\, x_n,\, v_{\iota_{\mathcal{K}}(x_n)}\big), \qquad
    u_{[n+1]} = u_{[n]} v_{\iota_{\mathcal{K}}(x_n)}, \\
    &u_{\mathcal{K},x} = \lim_{n\to\infty} u_{[n]}, \qquad
    t_{n+1} := t_{n} + \tau_{\iota_{\mathcal{K}}(x_n)}, \quad \forall n \geq 0.       
\end{align}

By construction, for all $n\geq 0$,
\[
0 < \min_{i\in\{0,\dots,N\}} \tau_i \;\leq\; t_{n+1}-t_n \;\leq\; \max_{i\in\{0,\dots,N\}}\tau_i =: \tau,
\]
so condition~\eqref{eq:sequence-condition-1} holds. Moreover, by induction one shows that for all $n\geq 1$,
\[
\phi(t_n,x,u_{\mathcal{K},x}) = x_n 
= \phi\big(\tau_{\iota_{\mathcal{K}}(x_{n-1})}, x_{n-1}, v_{\iota_{\mathcal{K}}(x_{n-1})}\big).
\]

We claim that if $x_n \in S$ then $x_{n+1}\in S$. Suppose first that $x_n \in S\setminus B_\varepsilon(x^*)$. By the covering condition~\eqref{eq:covering-support}, there exists $(x_i,r_i,v_i)\in\mathcal{K}$ such that $x_n\in B_{r_i}(x_i)$ and the verification condition~\eqref{eq:verification-condition} holds. In particular, by~\eqref{eq:verify-feasibility},
\begin{align}
\mathrm{sd}(x_{n+1},S) &= \mathrm{sd}(\phi(\tau_i,x_n,v_i),S) \\
&\leq \mathrm{sd}(\phi(\tau_i,x_i,v_i),S) + r_i e^{L\tau_i} \\
&\leq 0,
\end{align}
which implies $x_{n+1}\in S$.  

If instead $x_n\in B_\varepsilon(x^*)$, then either $\iota_{\mathcal{K}}(x_n)\neq 0$ and the above argument applies, or $\iota_{\mathcal{K}}(x_n)=0$, in which case we apply $v_0$ for time $\tau_0$. By the containment lemma applied to the ball $B_\varepsilon(x^*)$,
\begin{align}
\|x_{n+1}-x^*\| &= \|\phi(\tau_0,x_n,v_0)-x^*\| \\
&\leq \varepsilon + d(\phi(\tau_0,x_n,v_0),B_\varepsilon(x^*)) \\
&\leq \varepsilon + F_{B_\varepsilon(x^*)}\tau e^{L\tau_0} \\
&\leq \varepsilon(1+L\tau_0 e^{L\tau_0}), \label{eq:c-condition}
\end{align}
so by~\eqref{eq:covering-epsilon} and $\tau_0\leq\tau$ we conclude $x_{n+1}\in S$. Thus, $x_n\in S$ implies $x_{n+1}\in S$, i.e., condition~\eqref{eq:sequence-condition-2} holds.

\emph{Verification of \eqref{eq:sequence-condition-3}.}  
Let $\delta := \varepsilon(1+L\tau e^{L\tau})$ and $\bar n := \inf\{n : \|x_n-x^*\|\le\delta\}$.  
If $\iota_{\mathcal{K}}(x_n)=0$ and $\|x_n-x^*\|\le\varepsilon\le\delta$, then by~\eqref{eq:c-condition} we have $\|x_{n+1}-x^*\|\le\delta$.  

If $\iota_{\mathcal{K}}(x_n)=i\neq 0$, then from~\eqref{eq:verify-exp-decreasing} and $x_n\in B_{r_i}(x_i)$,
\begin{align}
e^{\alpha (t_{n+1}-t_n)}\|x_{n+1}-x^*\|
&\leq e^{\alpha\tau_i}\big(\|\phi(\tau_i,x_i,v_i)-x^*\| + r_i e^{L\tau_i}\big) \\
&\leq \|x_i-x^*\|-r_i \;\leq\; \|x_n-x^*\|,
\end{align}
which implies
\[
e^{\alpha(t_{n+1}-t_n)}(\|x_{n+1}-x^*\|-\delta) \;\leq\; \|x_n-x^*\|-\delta.
\]
If $\|x_n-x^*\|\le\delta$ ($n\ge\bar n$), then this inequality ensures $\|x_{n+1}-x^*\|\le\delta$.  
If $\|x_n-x^*\|>\delta$ ($n<\bar n$), iterating yields
\[
\|x_n-x^*\|-\delta \;\leq\; e^{-\alpha t_n}\big(\|x-x^*\|-\delta\big).
\]
Hence for $n\ge\bar n$, $\|x_n-x^*\|\le\delta$, while for $n<\bar n$ the excess above $\delta$ decays exponentially. This verifies condition~\eqref{eq:sequence-condition-3}.  

By Lemma~\ref{lem:RCLF-charac}, $V(x)=\|x-x^*\|$ is an R-CLF with rate $\alpha$ over $S$ and parameter $\delta=\varepsilon(1+L\tau e^{L\tau})$. Therefore, Theorem~\ref{thm:RCLF-stabilizability} implies that $x^*$ is practically exponentially stable on $S$ under $\pi_{\mathcal{K}}$, with constants
\[
\lambda=\alpha,\qquad 
K=e^{\alpha\tau}(1+L\tau e^{L\tau}),\qquad 
c=\varepsilon(1+L\tau e^{L\tau}).
\]
\end{proof}

\subsection{Existence and Sample Complexity of NCPs}
Theorem~\ref{thm:NPC} establishes that nonparametric chain policies can guarantee practical exponential stability of a region around an equilibrium point that is appropriately covered by data points from $\mathcal{K}$. However, it is a priori not clear how many data points are needed to construct such policy, or even whether such a policy exists.
The next result provides conditions for existence of Chain Policies as well as a bound on the sample complexity of such policies, i.e., the sizes of the assignment set $\mathcal{K}$ and alphabet $\mathcal{A}$ required to construct such policy.


\begin{theorem}[Existence and Sample Complexity of Chain Policies]\label{thm:sample-complexity}
Consider the control system~\eqref{eq:control-system} with equilibrium \(x^*\in\mathbb{R}^n\), and assume \(x^*\) is \(\lambda\)-exponentially stabilizable on \(\mathbb{R}^n\) with gain \(K > 0\).
Let \(S = B_R(x^*)\) with $R>0$, and choose $\varepsilon$ s.t. $R>\varepsilon>0$.
Fix any \(\alpha \in (0,\lambda)\) and choose
\begin{equation}\label{eq:rho}
\tau > \frac{\ln K}{\lambda - \alpha}, \quad 
L := L_{\mathcal{R}^{\tau}(S)}, \quad
\rho := \frac{1 - K e^{-(\lambda - \alpha)\tau}}{1 + e^{(L + \alpha)\tau}}.    
\end{equation}

Then there exists a nonparametric chain policy \(\pi_{\mathcal{K}}\) built from a finite assignment set of verification points \(\{(x_i,r_i)\}_{i=1}^N \subset S\) and associated controls \(\{v_i\}_{i=1}^N\) such that:
\begin{enumerate}[(i)]
\item \textbf{Practical exponential stability.} For every \(x \in S\), the induced closed loop satisfies
\[
\|\phi(t,x,u_{\mathcal{K},x}) - x^*\| \;\le\; C\,e^{-\alpha t}\|x - x^*\| + c, \quad \forall t \ge 0,
\]
with
$C = e^{\alpha\tau}\!\left(1 + L\tau e^{L\tau}\right)$  and $c = \varepsilon\!\left(1 + L\tau e^{L\tau}\right)$.
\item \textbf{Sample complexity.} The number \(N\) of covering centers and controls satisfies
\[
N \;=\; O\!\left(\left(\frac{3}{\rho}\right)^{\!d}\,\log\!\frac{R}{c}\right).
\]
\end{enumerate}
\end{theorem}

\begin{remark}[Performance--Complexity Trade-off]\label{rem:performance-complexity}
The definition of $\rho$ in~\eqref{eq:rho} reveals two contrasting regimes.
When $\lambda-\alpha$ is close to the lower bound $\ln K/\tau$, the numerator
$1 - K e^{-(\lambda-\alpha)\tau}$ approaches zero, so $\rho \approx 0$.
In this regime the guaranteed rate $\alpha$ is nearly as fast as the best
attainable $\lambda$, but the sample complexity bound
$O((3/\rho)^d)$ becomes extremely large. At the other extreme,
when $\alpha \ll \lambda$, the term $K e^{-(\lambda-\alpha)\tau}$ vanishes,
and $\rho$ approaches $1/(1+e^{L\tau})$. In this regime, far fewer
samples are needed, but the realized performance $\alpha$ is much slower than
the system’s intrinsic rate $\lambda$. 

Thus, $\rho$ quantifies the fundamental
trade-off: choosing $\alpha$ close to $\lambda$ yields strong performance at
the cost of high sample complexity, while smaller $\alpha$ reduces sample
requirements but sacrifices convergence speed.
\end{remark}

\ifthenelse{\boolean{arxiv}}{
\begin{proof}
Since $x^*$ is $\lambda$-exponentially stabilizable on $S$ with constant $K>0$, for each grid center $x_i \in S$ we can select a constant control $v_i$ such that 
\begin{equation}\label{eq:lambda-decay}
    \|\phi(t,x_i,v_i) - x^*\| \le K e^{-\lambda t} \|x_i - x^*\|, \quad \forall t \ge 0.
\end{equation}
We construct the nonparametric chain policy $\pi_{\mathcal{D}}$ by covering $S$ with finitely many balls $B_{r_i}(x_i)$ and assigning to each $x_i$ the control $v_i$ above. In our construction, we will assume $\|\cdot\|$ is the infinity norm, i.e., $\|x\| = \max_{j=1,\dots,n} |x_j|$.

\noindent\textbf{Step 1: Choice of radii.}
Fix $\alpha \in (0,\lambda)$ and $\tau > \frac{\ln K}{\lambda - \alpha}$.  
Let $L := L_{\mathcal{R}^\tau(S)}$ and set
\[
\rho := \frac{1 - K e^{-(\lambda - \alpha)\tau}}{1 + e^{(L + \alpha)\tau}} > 0,
\quad
r_i := \rho \,\|x_i - x^*\|.
\]
By construction $r_i > 0$.  From \eqref{eq:lambda-decay}, for $t = \tau$ we have
\[
K e^{-(\lambda - \alpha)\tau} \|x_i - x^*\| + r_i e^{(L+\alpha)\tau}
= \|x_i - x^*\| - r_i,
\]
which implies the verification condition
\begin{align*}
    \min_{t \in (0,\tau]} & e^{\alpha t} \big(\|\phi(t,x_i,v_i) - x^*\| + r_i e^{Lt}\big) \leq\\ 
    &\le e^{\alpha\tau}\!\left(\|\phi(\tau,x_i,v_i) - x^*\| + r_i e^{L\tau}\right) \\
    &\le \|x_i - x^*\| - r_i.
\end{align*}
Hence each $(x_i,r_i)$ satisfies~\eqref{eq:verify-exp-decreasing}.

\noindent\textbf{Step 2: Covering number bound.}
We cover the annular region $B_R(x^*) \setminus B_{\varepsilon}(x^*)$ by $n$ concentric annuli $A_1,\dots,A_n$ of thickness $R_i = 3^{i-1}\varepsilon$.  
Since the total radial width is $R-\varepsilon$,  
\begin{align*}
R - \varepsilon \le \sum_{i=1}^n 3^{i-1} \varepsilon
= \varepsilon\,\frac{3^n - 1}{2}
\quad\Rightarrow\quad\\
n = \left\lceil \log_3\!\left(\frac{2R}{\varepsilon} - 1\right) \right\rceil
= O\!\left(\log\frac{R}{\varepsilon}\right).
\end{align*}

Each $A_i$ is initially partitioned into $3^d - 1$ hypercubes of side $R_i$.  
For any $x \in A_i$, $\|x - x^*\| \ge 3^{i-1} \varepsilon$ and $r_i \le \rho\|x - x^*\|$.  
We refine each hypercube by successive splits into $3^d$ subcubes until the side length is at most $\rho\|x - x^*\|$.  
This requires
\[
3^{m} \ge \frac{1}{\rho}
\quad\Rightarrow\quad
m = \lceil \log_3(1/\rho) \rceil
\]
splits, producing at most
\[
N_{\text{annulus}} = (3^d - 1)\,3^{dm} = O\!\left( \left(\frac{3}{\rho}\right)^d \right)
\]
points per annulus.

\noindent\textbf{Step 3: Total number of points.}
Multiplying by the number of annuli,
\[
N = n \cdot N_{\text{annulus}}
= O\!\left( \left(\frac{3}{\rho}\right)^d \log\frac{R}{\varepsilon} \right).
\]
Finally, since $c=\varepsilon(1+L\tau e^{L\tau})$, this yields the claimed sample complexity bound.
\end{proof}
}{
\noindent \textit{Proof.}~The proof is omitted due to page limits. \hfill \qed
}

\subsection{Incremental Learning of NCPs}

The performance--complexity trade-off discussed in Remark~\ref{rem:performance-complexity}, together with the existence and sample complexity result of Theorem~\ref{thm:sample-complexity}, suggests a
practical methodology for progressively improving performance. By actively sampling trajectories more finely and refining the covering set $\mathcal{K}$, one can construct NCPs that certify larger rates $\alpha$ (by reducing the effective radius $r$), thereby reducing the gap between realized performance and the best attainable rate $\lambda$. In other words, performance can be systematically enhanced by enriching the assignment set with additional verified points, which readily enables incremental learning.

Beyond improving rates, another key feature of NCPs is their ability to incrementally \emph{expand the certified region}. The next result formalizes this incremental learning
property: previously verified assignments can be reused together with new ones to grow the domain over which stability is guaranteed.


\begin{theorem}[Incremental Learning of $\mathcal{K}$] \label{thm:incremental-covering}
Consider an equilibrium point $x^*\in \mathbb{R}^n$ of \eqref{eq:control-system} and a set $S\subseteq \mathbb{R}^n$ satisfying $x^*\in \mathrm{int}(S)$. Let $\pi_\mathcal{K}$ be a nonparametric policy with assignment set $\mathcal{K}=\{(x_i,r_i,v_i)\}_{i=1}^N$  and default control $v_0\in\mathcal{A}$ satisfying properties \emph{(i)–(iii)} of Theorem~\ref{thm:NPC} with parameters $\alpha,\,\delta,\,\tau,\,L,\,\varepsilon$. 
Take $x_j\in \mathbb{R}^n\setminus S$, $r_j>0$, and $v_j\in\mathcal{U}^{[0,\tau_j)}$ s.t. $B_{r_j}(x_j)\cup S=\emptyset$. 
Define the enlarged set $S' := S \cup B_{r_j}(x_j)$, and let $L_j = L_{\mathcal{R}^{\tau_j}(B_{r_j}(x_j))}$, and $L'=\max\{L_j,L\}$.

Whenever the following conditions are satisfied:
\begin{enumerate}[(1)]
    \item \textbf{Feasibility of $(x_j,r_j,v_j)$:} The 3-tuple $(x_j,r_j,v_j)$ with $v_j:[0,\tau_j)\rightarrow U$ satisfy 
    \begin{align}
        \mathrm{sd}(\phi(\tau_j,x_j, v_j),S)+r_j e^{L_j\tau_j}&\leq 0.
    \end{align}
    \item \textbf{Either of the following holds}:
    \begin{enumerate}[(a)]
        \item \textbf{Direct verification at $(x_j,r_j,v_j)$:} The tuple satisfies decrease condition: 
        \begin{align}\label{eq:verify-decrease-x_j}
            e^{\alpha \tau_j}\bigl(\|\phi(\tau_j,x_j, v_j)\!-\!x^*\|\!+\!r_j e^{L\tau_j}\bigr)\!\leq\! \|x_j \!-\! x^*\|\!-\!r_j.
        \end{align}
        Set $\alpha'=\alpha$, $\tau'=\max\{\tau,\tau_j\}$, $\delta'=\delta$.
        \item \textbf{Bootstrapping:} There is $\mathcal{\hat K}\subseteq \mathcal{K}$, such that 
        \begin{enumerate}[(i)]
            \item $B_{r_je^{L_j\tau_j}}(\phi(\tau_j,x_j,v_j))\subseteq \mathrm{Supp}(\mathcal{\hat K})$
            \item There is $\alpha'<\alpha$ such that 
        \[
        \max_{(x_i,r_i,v_i)\in\mathcal{\hat K}} \frac{e^{-(\alpha -  \alpha')\tau_i}}{e^{-\alpha \tau_j}}  
        \frac{\|x_{i}-x^*\|+r_{i}}{\|x_j-x^*\|-r_j} \leq 1.
        \]
        \end{enumerate}
        Set $\tau'= \tau +\tau_j$, $\delta'=\delta$.
    \end{enumerate}
\end{enumerate}
Then the augmented assignment set $\mathcal{K}' := \mathcal{K}\,\cup\,\{(x_j,r_j,v_j)\}$
and the default control $v_0$ induce a policy $\pi_{\mathcal{K}'}$ that practically exponentially stabilizes $x^*$ over $S'=S\cup B_{r_j}(x_j)$ with 
\[
\lambda'=\alpha', \quad 
K'= e^{\alpha'\tau'}(1+L'\tau e^{L'\tau'}), \quad
c'=\varepsilon(1+L\tau e^{L\tau}).
\]
\end{theorem}

\begin{proof}
We first note that since $\mathcal{K}$ and $v_0$ satisfy  Theorem~\ref{thm:RCLF-stabilizability} (i)–(iii) on $S$, and  $S \cap B_{r_j}(x_j) = \emptyset$, any initial state $x \in S$ under  $\pi_{\mathcal{K}'}$ will only trigger assignments from $\mathcal{K}$ or the  default control $v_0$. By Theorem~\ref{thm:NPC}, this ensures that for every  $x \in S$ there exists a sequence of times $\{t_n\}$ satisfying the conditions 
of Lemma~\ref{lem:RCLF-charac}, along which \eqref{eq:sequence-condition} is satisfied for  $V(x)=\|x-x^*\|$. Hence, whenever $x\in S\subset S'$, the trajectory  $\phi(t,x,u_{\mathcal{K}',x})$ is practically exponentially stable with  parameters $\lambda=\alpha$, $K=e^{\alpha\tau}(1+L\tau e^{L\tau})$, and  $c=\varepsilon(1+L\tau e^{L\tau})$. Moreover, since $\alpha \leq \alpha'$,  $\tau'\geq \tau$, and $L'\geq L$, the same trajectory also satisfies the  practical exponential stability bound with the updated constants $\lambda'$, $K'$, and $c'$ as stated in the theorem.

We next consider the case $x \in S'\setminus S$. Since $S \cap B_{r_j}(x_j) = \emptyset$, it follows that $\iota_{\mathcal{K}'}(x) = j$, so the first control applied by $\pi_{\mathcal{K}'}$ is $v_j$.
We will use the sequence \eqref{eq:npc-sequence-K} induced by $\pi_{\mathcal{K'}}$ to build a sequence of times $\{t_n\}_{n\in\mathbb{N}}$ satisfying the properties in \eqref{eq:sequence-condition} of Lemma \ref{lem:RCLF-charac} for the control signal $u_{\mathcal{K'},x}$. The result then follows from Theorem \ref{thm:RCLF-stabilizability}.  We will choose $\delta'$ and $\alpha'$ in \eqref{eq:sequence-condition} later on.
Recall that $\pi_\mathcal{K}$ satisfies \eqref{eq:sequence-condition} for some $\alpha$ and $\delta = \varepsilon(1+L\tau e^{L\tau})$.
To simplify notation, we use $u$ to refer to $u_{\mathcal{K'},x}$.

We choose $x_0=x$ and $t_0=0$. We will select $t_1$ differently, depending on which clause of condition (2) in the theorem's hypothesis hold.

\noindent
\textbf{Case (2a).} We choose $t_1=\tau_j$, and accordingly $x_1 = \phi(\tau_j,x,v_j)=\phi(t_1,x,u)$.
By condition (1) of the theorem, we have
\begin{align*}
\mathrm{sd}\!\left(\phi(\tau_j,x,v_j),S\right) 
&\leq \mathrm{sd}\!\left(\phi(\tau_j,x_j,v_j),S\right) 
      + r_j e^{L_j \tau_j} \leq 0,
\end{align*}
which implies $\phi(\tau_j,x,v_j)\in S$. This ensures $\phi(t_1,x,u)=\phi(\tau_j,x,v_j)\in S\subset S'$, i.e., \eqref{eq:sequence-condition-2} for the set  $S'$.
It further follows from \eqref{eq:verify-decrease-x_j} by a similar argument to Theorem \ref{thm:NPC}, that 
\[
\|\phi(t_1,x,v_j)-x^*\| \;\leq\; e^{-\alpha t_1}\,\|x-x^*\|,
\]
so \eqref{eq:sequence-condition-3} holds for $\alpha$ and any value of $\delta$.

\noindent\textbf{Case (2b).} As in case (2a) after choosing $v_j$, $\phi(\tau_j,x,v_j)\in S$. In fact, by (2b.i), $\phi(\tau_j,x,v_j)\in \mathrm{Supp}(\hat{\mathcal K})\subseteq S$. Let $(x_i,r_i,v_i)\in\hat{\mathcal K}$ s.t. $i=\iota_{\mathcal{K}}(\phi(\tau_j,x,v_j))$. 
We choose
\begin{align}
t_1&:=\tau_j+\tau_i,\\ 
x_1&:=\phi(\tau_j+\tau_i,x,v_jv_i)=\phi(\tau_i,\phi(\tau_j,x,v_j),v_i).
\end{align}
Since \eqref{eq:verification-condition} holds for $(x_i,r_i,v_i)\in\mathcal{K}$, $x_1\in S$, and therefore we have  $\phi(t_1,x,u)=x_1\in S\subset S'$; hence  \eqref{eq:sequence-condition-2} holds for $t_1$ and $S'$.

It remains to ensure the decrease in \eqref{eq:sequence-condition-3} at $t_1$ with some 
$\alpha'<\alpha$ and $\delta'=\delta$. We thus consider
\begin{align}
e^{\alpha't_1}\|&\phi(t_1,x,u)-x^*\|=\\
&=e^{\alpha'(\tau_j+\tau_i)}\|\phi(\tau_j+\tau_i,x,v_j v_i)-x^*\|\\
&=e^{\alpha'(\tau_j+\tau_i)}\|\phi(\tau_i,\phi(\tau_j,x,v_j),v_i)-x^*\|\\
&\leq e^{\alpha'(\tau_j+\tau_i)} e^{-\alpha\tau_i}\|\phi(\tau_j,x,v_j)-x^*\|\\
&= \frac{e^{-(\alpha-\alpha')\tau_i}}{e^{-\alpha'\tau_j}} \left(\frac{\|\phi(\tau_j,x,v_j)-x^*\|}{\|x-x^*\|}\right)\|x-x^*\|\\
&\leq \frac{e^{-(\alpha-\alpha')\tau_i}}{e^{-\alpha'\tau_j}}\left(\frac{\|x_i-x^*\|+r_i}{\|x_j-x^*\|-r_j}\right)\|x-x^*\|\\
&\leq \|x-x^*\|\,,
\end{align}
where step one follows from definition of $t_1$ and $u$, two from the group property of $\phi$, three from \eqref{eq:verify-exp-decreasing} on $(x_i,r_i,v_i)$, and the final step from the bootstrapping condition (2b.ii).

In both cases, after $t_1$ the subsequent points are generated by $\mathcal{K}$, 
so the sequence $\{t_n\}$ continues to satisfy 
\eqref{eq:sequence-condition-2}–\eqref{eq:sequence-condition-3} by 
Theorem~\ref{thm:NPC}. Thus $V(x)=\|x-x^*\|$ is an R-CLF on $S'$, and by 
Theorem~\ref{thm:RCLF-stabilizability}, $\pi_{\mathcal{K}'}$ renders $x^*$ 
practically exponentially stable on $S'$ with the claimed constants.
\end{proof}


    

 \section{Numerical Experiments
}\label{ct-numerical}

Using the sufficiency of NCPs derived in Theorem~\ref{thm:NPC} along with the grid construction of Theorem~\ref{thm:sample-complexity}, we next introduce an algorithmic methodology to design NCPs to stabilize a given region. The algorithmic flow is as follows:

\begin{enumerate}[i.]
    \item Given a region $S \subset \mathbb{R}^n$, select a desired convergence rate $\alpha > 0$, select a $\tau_{\max}$ to upper-bound $\tau_i$, and determine the one-sided Lipschitz constant $L$ for the underlying dynamics across $\mathcal{R}^{\tau_{\max}}(S)$\footnote{Achieved through estimating the reachable tube by simulating samples along the boundary and adding precision-correction terms, and then sampling points in that region for OSL while adding precision-correction terms again.}.
    \item Create a grid of points and radii $G = \{(x_i, r_i)\}$ covering the region $S$ with increasingly large radii per annulus according to Theorem~\ref{thm:sample-complexity}.
    \item For each $x_i$, derive controls $v_i$ for $\tau_{\max}$ time (for example, using sampling methods akin to Model-Predictive Path Integrals (MPPI)~\cite{MPPI1, MPPI2}).
    \item For each $(x_i, r_i,v_i)$, if condition \eqref{eq:verify-exp-decreasing} is satisfied for some $\tau_i\leq\tau_{\max}$, store the largest $\tau_i$ that achieves equality.
    \item Otherwise, split the ball $B_{r_i}(g_i)$ into $3^d$ smaller balls. 
    \item Repeat until a control is found for $\alpha$-exponential decrease with slack of at least each ball's radius, or until a pre-defined maximum number of splits is achieved.
    \item For balls which still fail the conditions, apply condition 2b of Theorem~\ref{thm:incremental-covering} to leverage previously derived controls.
    \item Trim down the verified region only to those trajectories which satisfy \eqref{eq:verify-feasibility}, save verified $\alpha_i \geq \alpha$ for each cell.
\end{enumerate}

With this algorithmic method, we present a number of case studies in different classic control stabilization problems, which will demonstrate useful features of NCPs. 



\subsection{Unicycle}
\begin{figure}[!h]
  \centering

    \begin{subfigure}[b]{0.4925\columnwidth}
    \includegraphics[width=\linewidth]{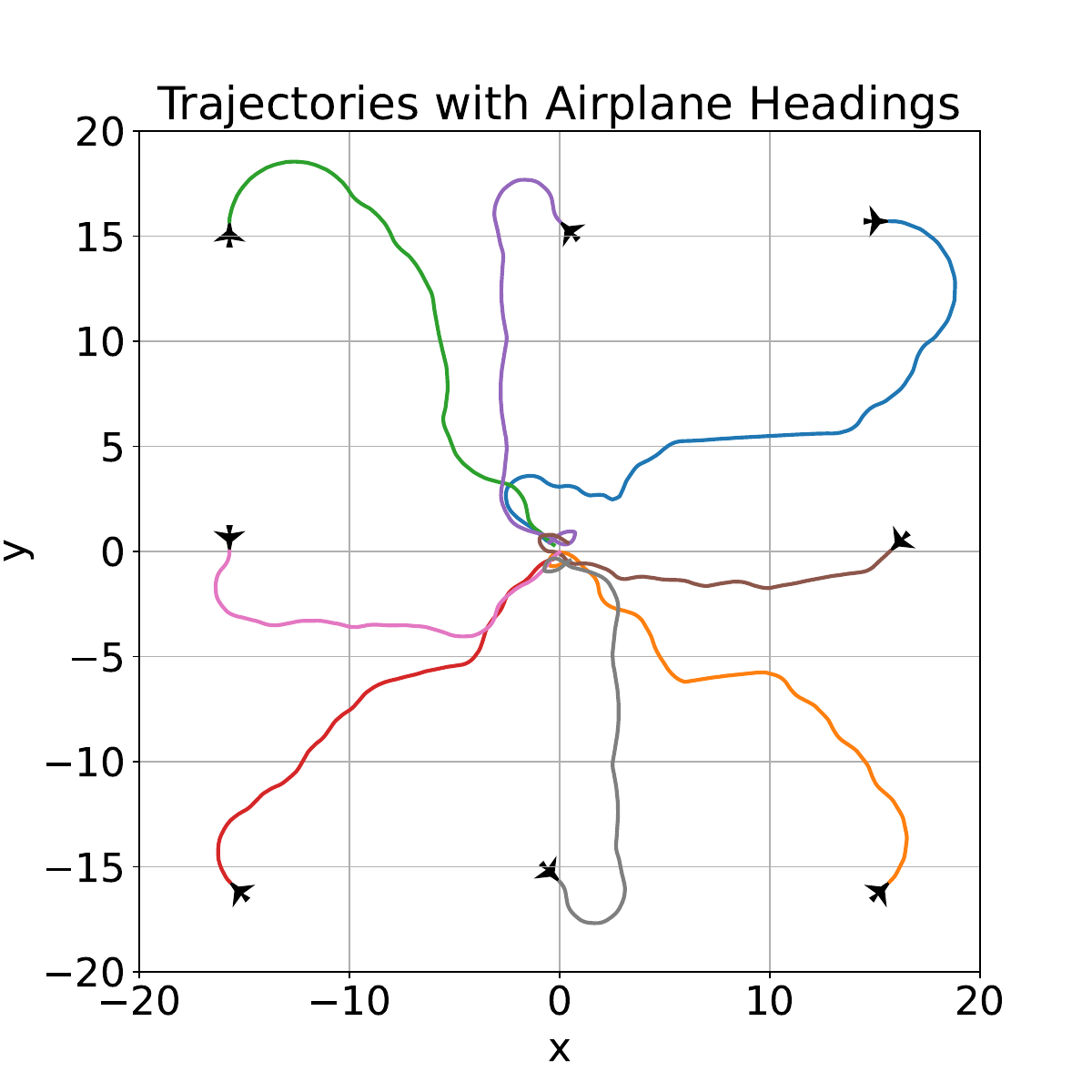}
    \caption{NCP trained with $V_1$.}
  \end{subfigure}
  \hfill
  \begin{subfigure}[b]{0.4925\columnwidth}
    \includegraphics[width=\linewidth]{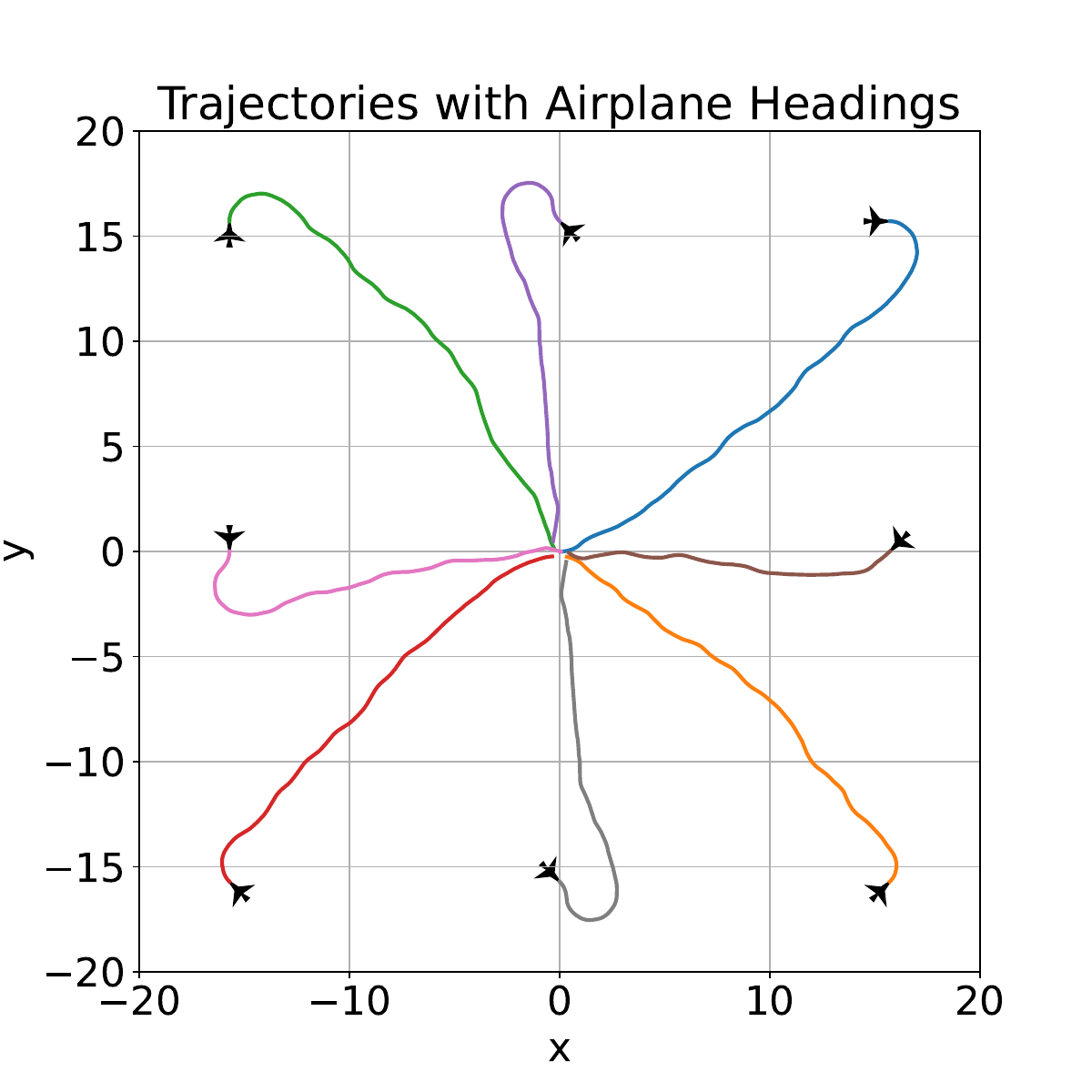}
    \caption{NCP trained with $V_2$.}
  \end{subfigure}
  
  \vspace{0.1em}

  \begin{subfigure}[b]{0.4925\columnwidth}
    \includegraphics[width=\linewidth]{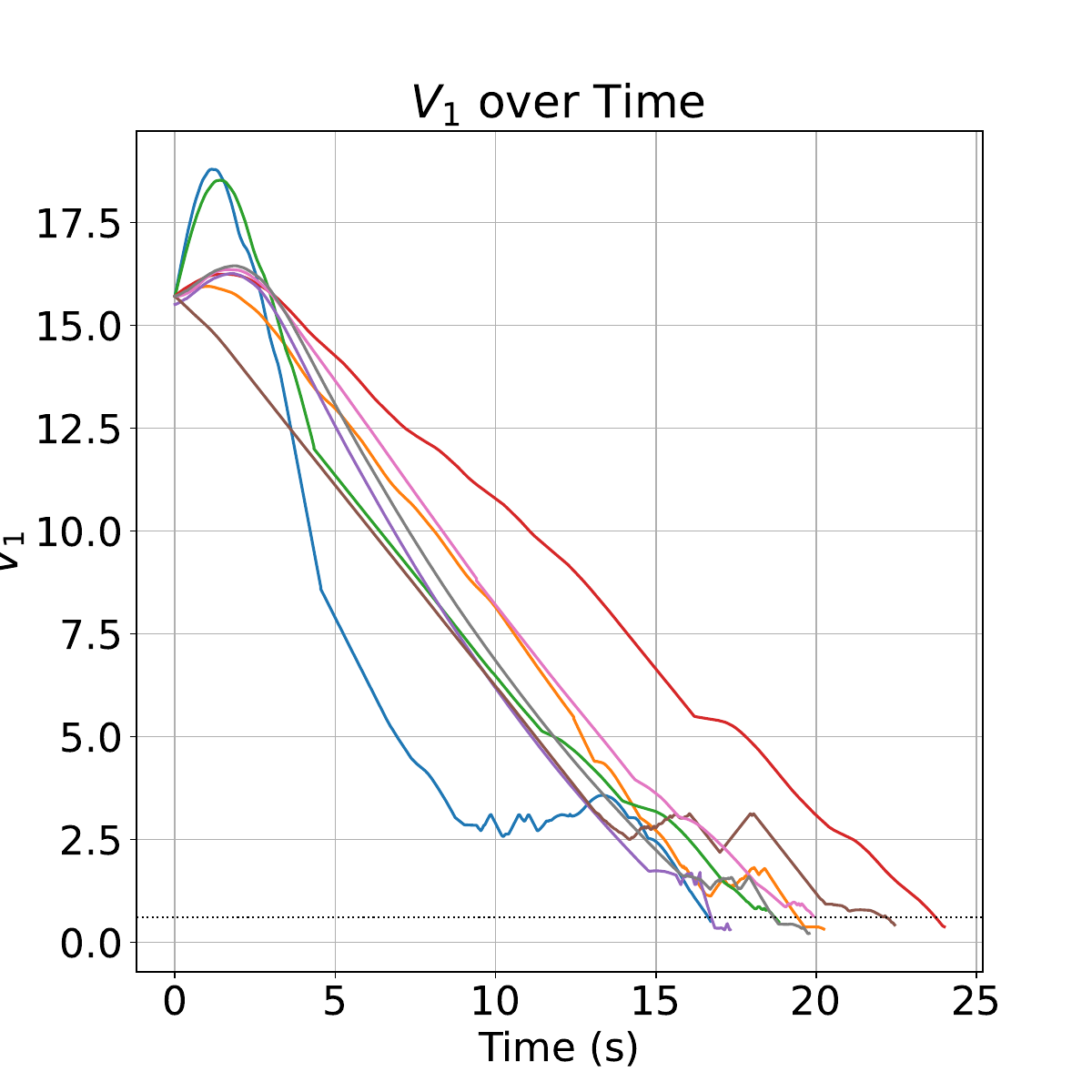}
    \caption{$V_1(\phi(t,x,u)).$}
  \end{subfigure}
  \hfill
  \begin{subfigure}[b]{0.4925\columnwidth}
    \includegraphics[width=\linewidth]{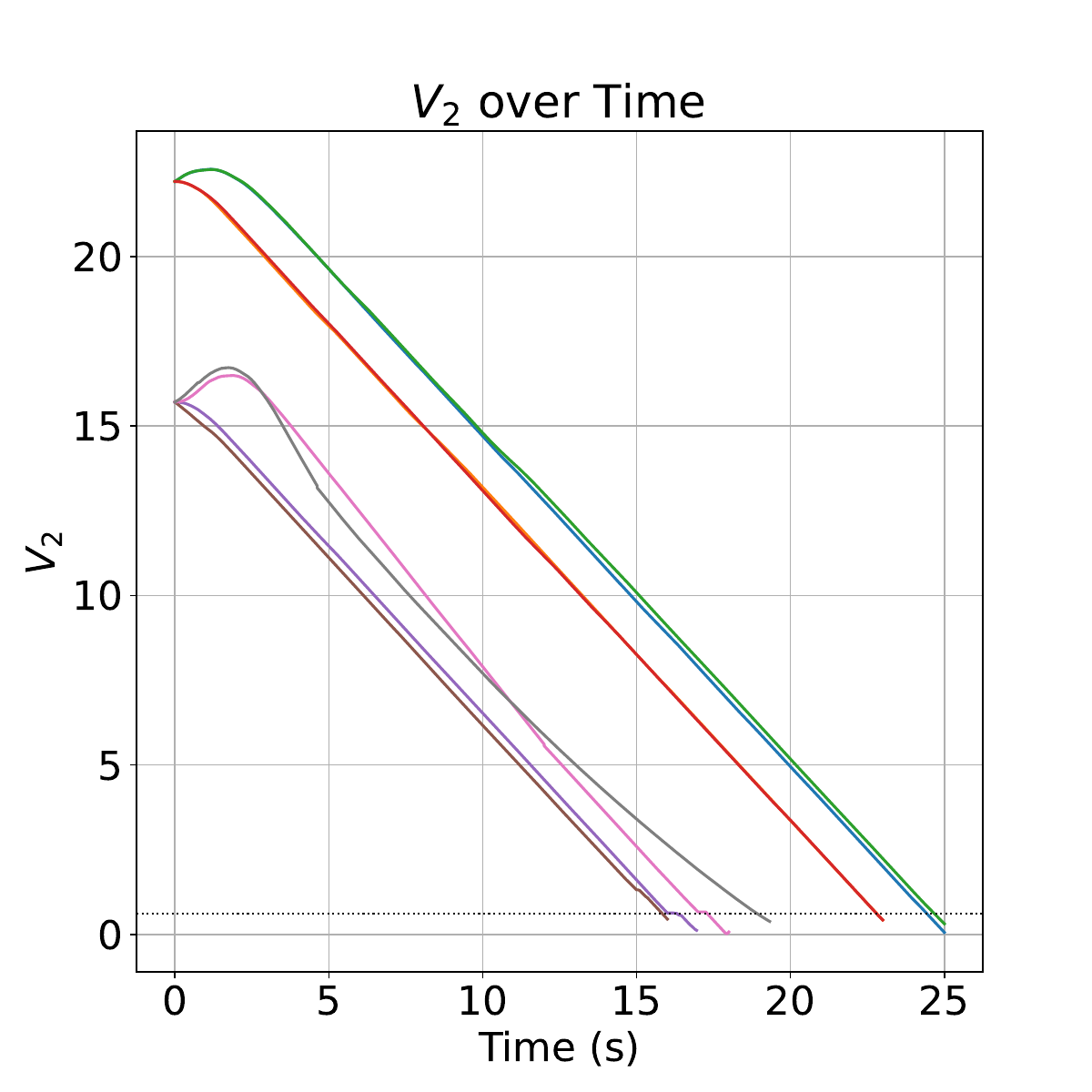}
    \caption{$V_2(\phi(t,x,u))$.}
  \end{subfigure}

  \caption{\textbf{Trajectories of Unicycle NCP.} Phase plots of $(x,y)$ for eight evenly distributed points. The black icons depict the initial facing of the unicycle. Plot (a) contains trajectories from NCP trained to minimize $V_1$, which results in sharp turns, while (b) is trained to minimize $V_2$, which results in softer turns and smoother overall behavior. Plots (c) and (d) show the development of $V_1$ and $V_2$ over time respectively. Both converge exponentially to the equilibrium, with at least$\alpha = 0.01$. We have $\tau_{\max} = 5, \varepsilon = 0.01, L = 1$ and $c = \varepsilon(1+L\tau_{\max} e^{L \tau_{\max}}) \simeq 0.613$, represented by the dotted line.}
  \label{fig:UniCycle}
\end{figure}

Consider the unicycle model moving in the plane, parametrized by its $x$ position, $y$ position, and angle of vehicle $\theta$ with respect to the $x$-axis, with two control inputs - velocity $v$ and angular velocity $\omega$. The dynamics are given by 
$$\begin{bmatrix} \dot{x} \\ \dot{y} \\ \dot{\theta} \end{bmatrix} = \begin{bmatrix}
    v \cos (\theta) \\ v \sin (\theta) \\ \omega
\end{bmatrix}.$$
We bound $u \in [0,1]$ and $\omega \in [-1, 1]$, and run the method to derive NCPs for two different norms, being $V_1 = \max\{|x|, |y|, |\theta|\}$ and $V_2 = \sqrt{x^2 + y^2 + 0.01\theta^2}$. For either choice of norm (simpler or tied to the classic reward function), the method quickly stabilized the entire region of $(x, y) \in [-20, 20]^2, \theta \in (-\pi, \pi]$, see Figure~\ref{fig:UniCycle}. 

\begin{figure}[!h]
  \centering
    \begin{subfigure}[b]{0.4925\columnwidth}
    \includegraphics[width=\linewidth]{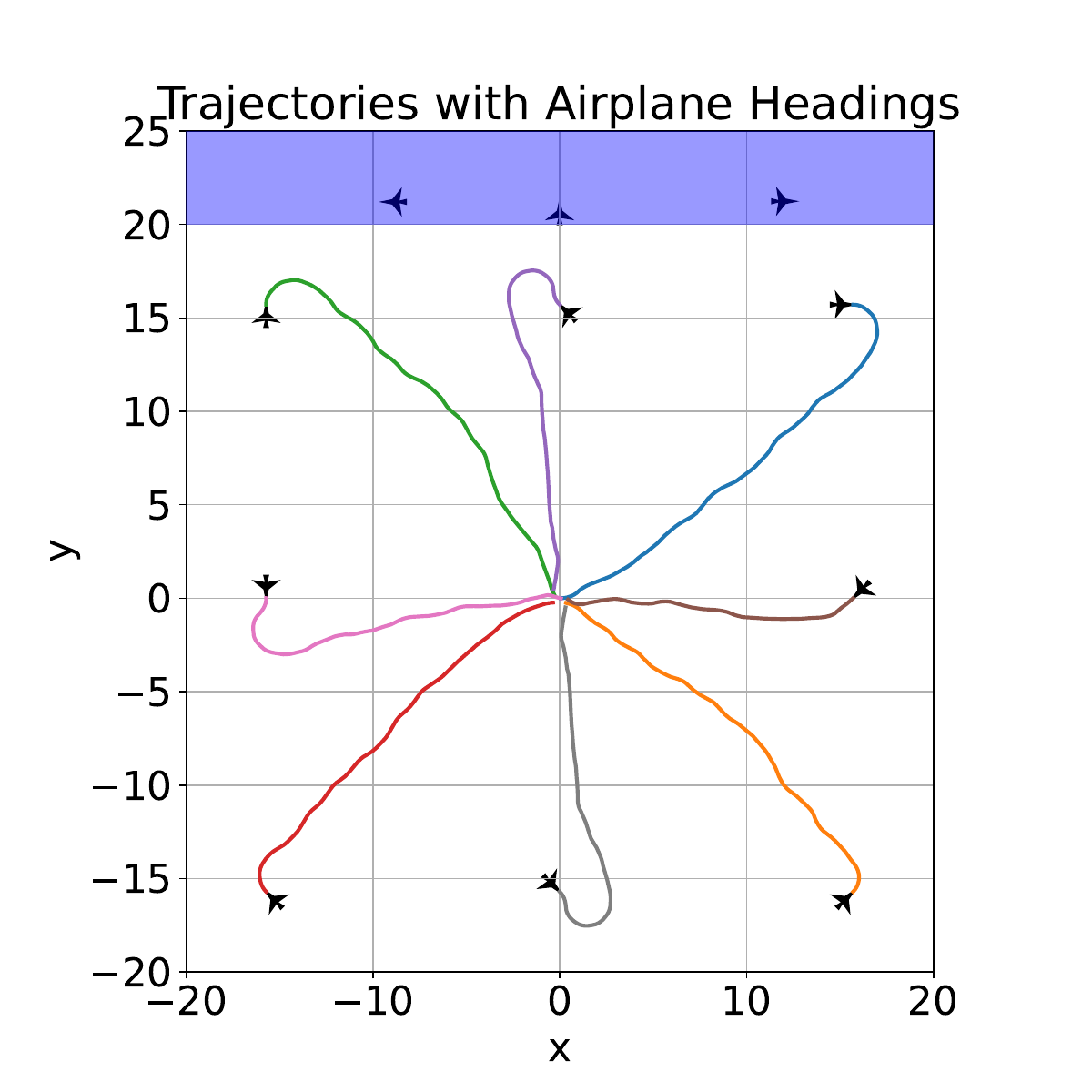}
    \caption{Before learning}
  \end{subfigure}
  \hfill
  \begin{subfigure}[b]{0.4925\columnwidth}
    \includegraphics[width=\linewidth]{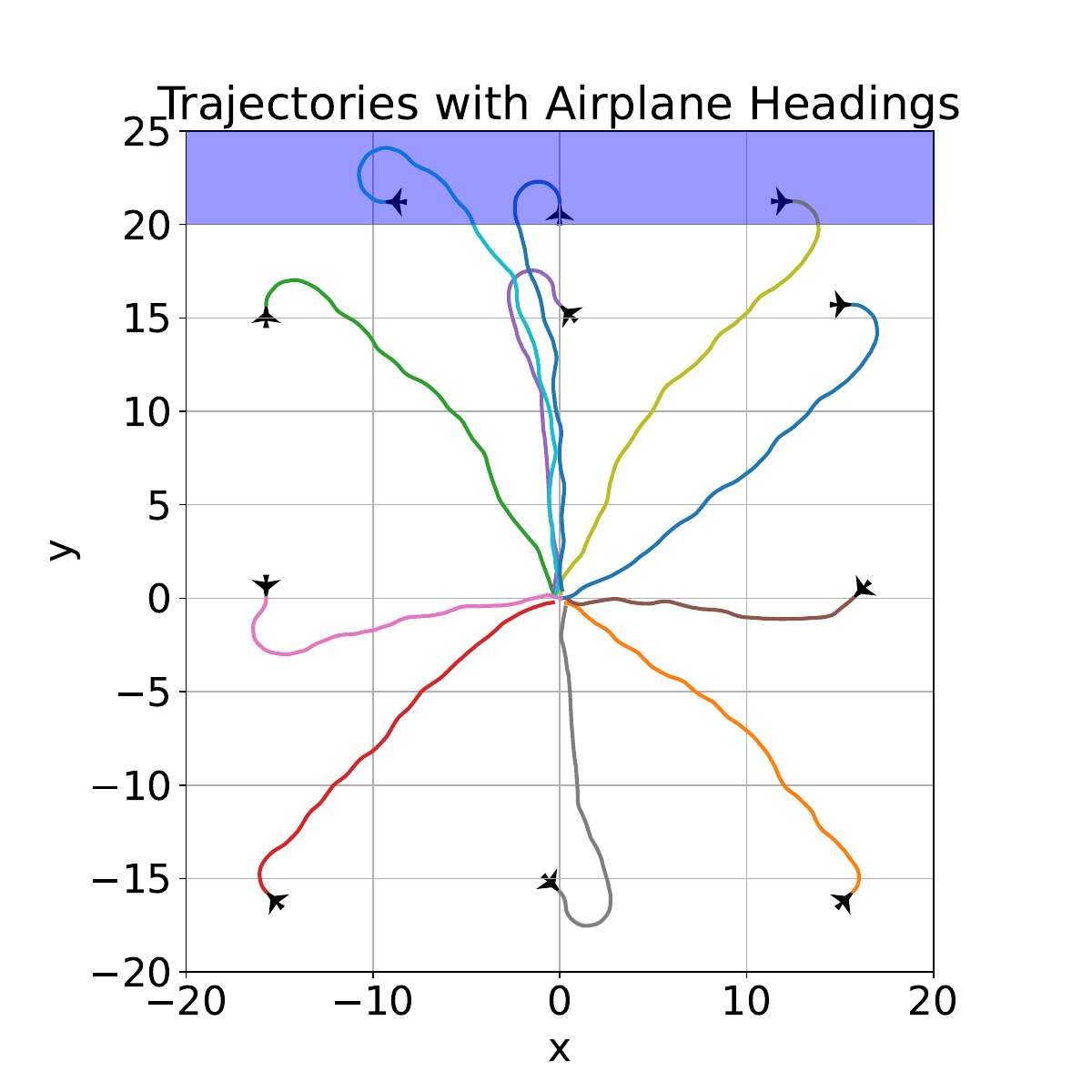}
    \caption{After learning.}
  \end{subfigure}

  \caption{\textbf{Incremental Learning of Unicycle Policy.} Extending the state space from the previously learned region in the $y$-direction. Subfigure (a) contains the phase plot before learning, while subfigure (b) contains the phase plot after. The new region is learned without forgetting, such that parts of the trajectory in the old region use previously designed controls.}
  \label{fig:UniCycleExpand}
\end{figure}

To demonstrate the incremental growth capabilities of NCP, we do two stages of learning. After learning a control for the previous region, i.e., $(x, y,\theta) \in [-20, 20]^2\times(-\pi,\pi]$, we expand the state space to include values in the region $(x,y,\theta) \in [-20, 20] \times [20, 25] \times (-\pi, \pi]$. 
Trajectories fragments starting in the formerly verified region retain the same behavior, while the new behavior (for initial values in the new region) are depicted in Figure~\ref{fig:UniCycleExpand}.

\subsection{Inverted Pendulum}

We next analyze the utility of the NCPs on the inverted pendulum. The system consists of a mass $m$ attached at the end of a rigid pendulum of length $l$, pivoting freely about a fixed point. The dynamics are governed by the torque around the pivot due to both gravity and an external control input. Denoting by $\theta$ the angle of the pendulum measured from the vertical (with $\theta=0$ corresponding to the inverted equilibrium), the equation of motion is given by
\begin{equation}
    ml^2\ddot{\theta}(t) = mgl\sin(\theta(t)) + u(t),
\end{equation}
where $g$ is the gravitational acceleration, and $u(t)$ is the external control torque applied at the pivot.

\begin{figure}[!h]
  \centering

    \begin{subfigure}[b]{0.4925\columnwidth}
    \includegraphics[width=\linewidth]{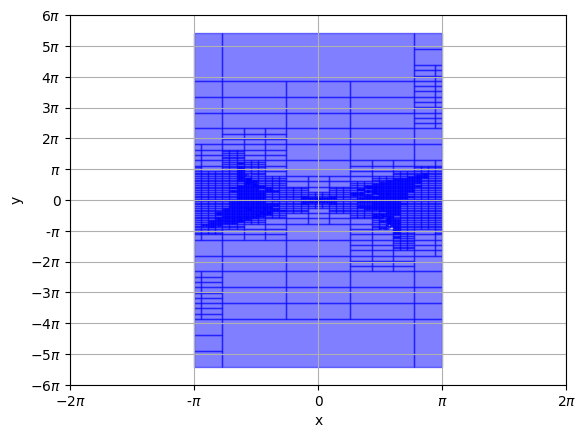}
    \caption{Base NCP verification.}
  \end{subfigure}
  \hfill
    \begin{subfigure}[b]{0.4925\columnwidth}
    \includegraphics[width=\linewidth]{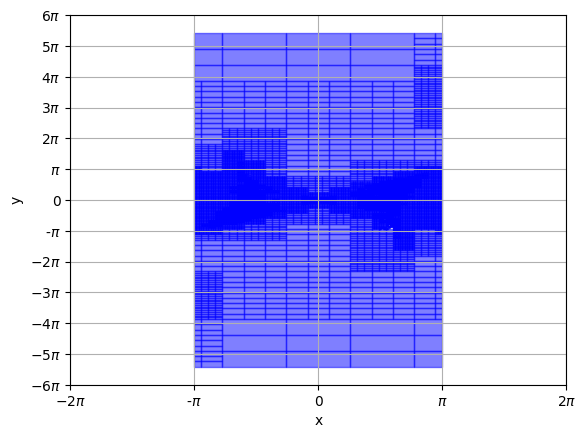}
    \caption{NCP refinement.}
  \end{subfigure}

  \vspace{0.1em}

  \begin{subfigure}[b]{0.4925\columnwidth}
    \includegraphics[width=\linewidth]{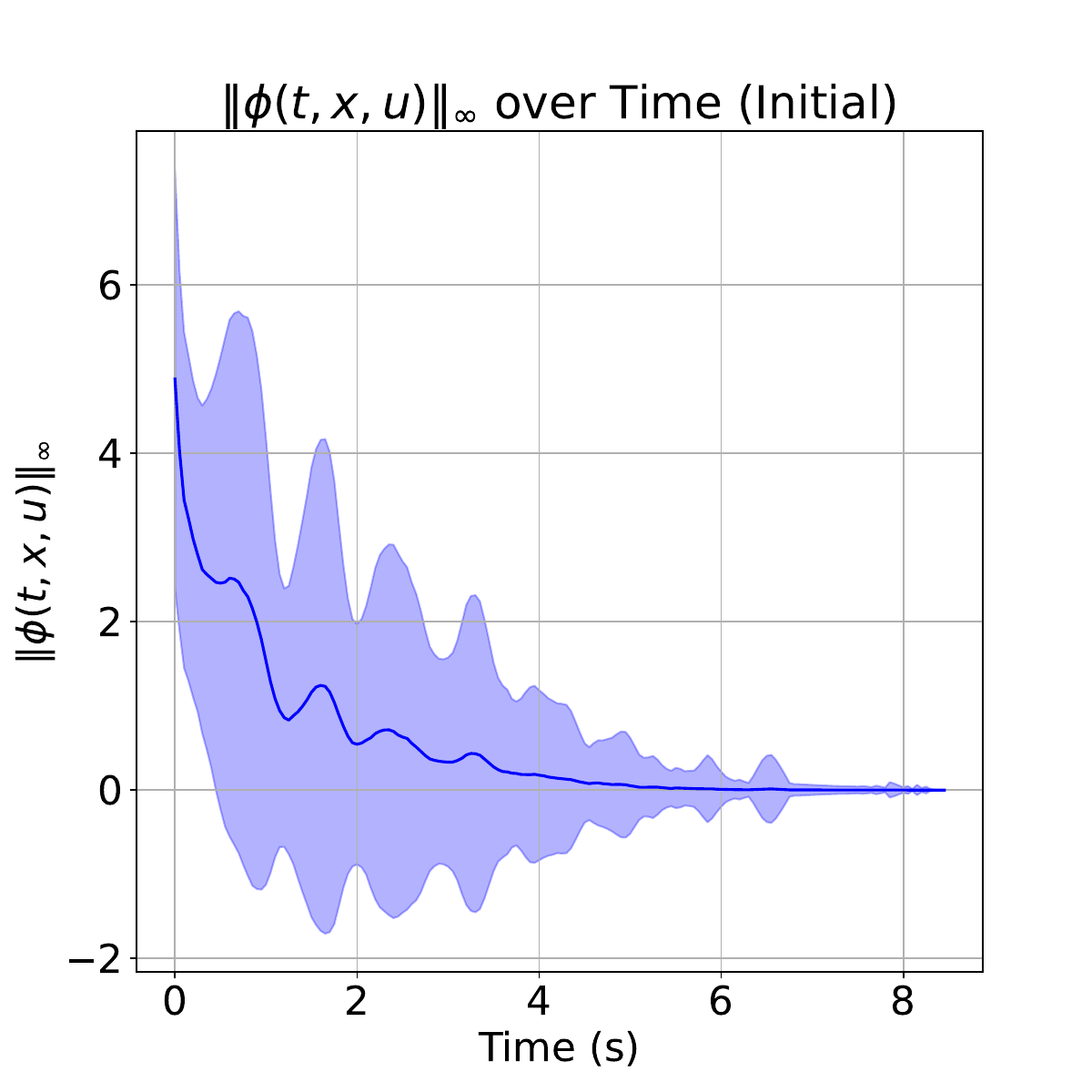}
    \caption{$\|\phi(t,x,u))\|_\infty$, initial.}
  \end{subfigure}
  \hfill
  \begin{subfigure}[b]{0.4925\columnwidth}
    \includegraphics[width=\linewidth]{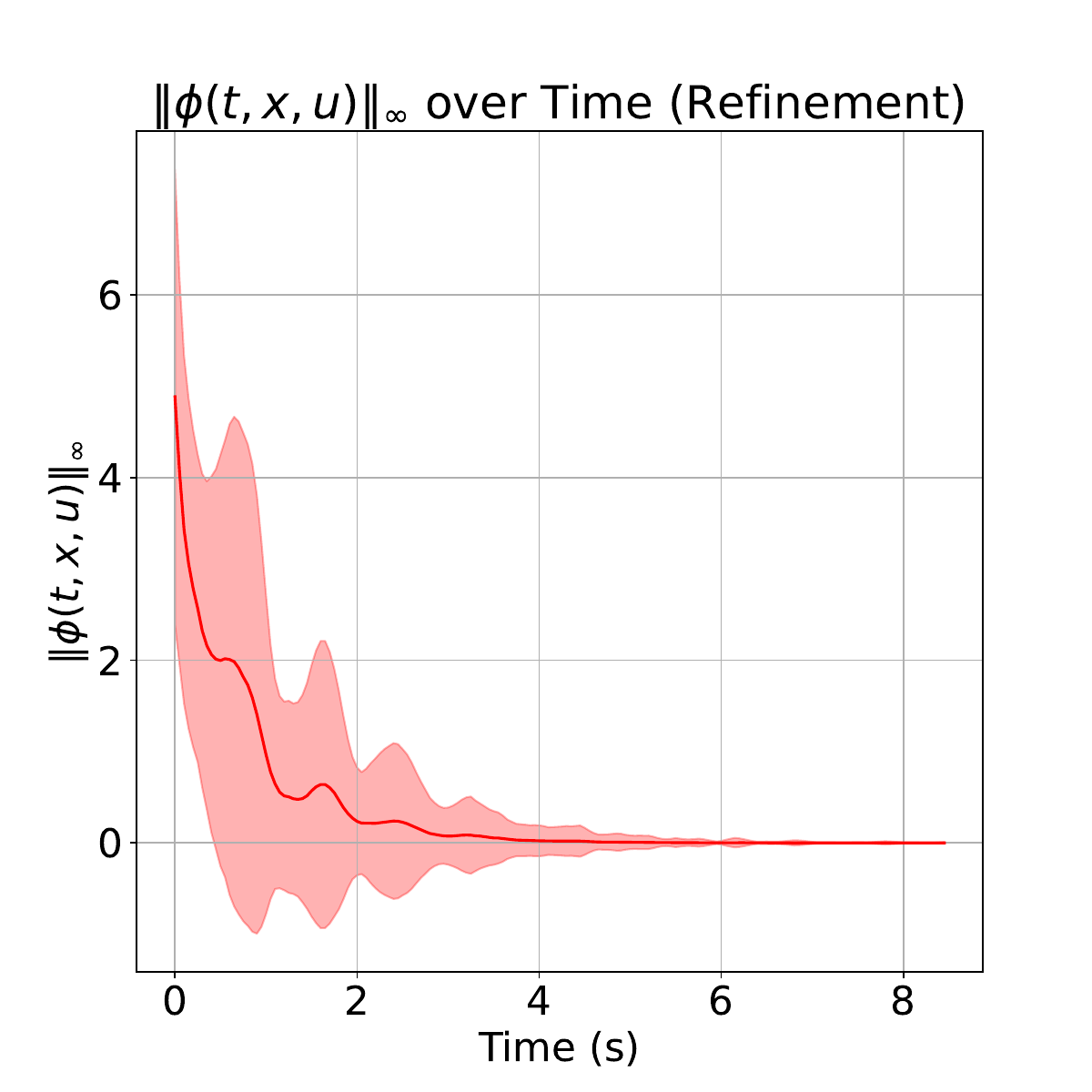}
    \caption{$\|\phi(t,x,u))\|_\infty$, refinement.}
  \end{subfigure}

  \caption{\textbf{Additional Data Refinement Facilitates Improved NCP Performance.} Plot (a) contains the balls used to verify the region $(\theta, \dot{\theta)} \in (-\pi,\pi] \times [-5\pi, 5\pi]$ for the inverted pendulum. Plot (b) is a refinement of plot (a), wherein all balls were split once more and re-verified. The minimum verified rate of convergence for trajectories $\alpha$ goes from $0.003$ to $0.0145$, and the average verified $\alpha$ goes from $1.815$ to $3.149$. Plot (c) demonstrates the average norm over time of 400 sample trajectories under each schema.
  We have $\tau_{\max} = 1.5$, $\varepsilon = 0.01$, $L = 1$, and $c = \varepsilon(1+L\tau_{\max}e^{L\tau_{\max}}) \simeq 0.072$. 
  }
  \label{fig:Refinement}
\end{figure}

Figure~\ref{fig:Refinement} demonstrates the refinement capabilities of NCPs, such that by adding data (simulated by splitting all balls once), the rate of convergence achieved is significantly increased.

\section{Conclusions}\label{ct-conclusions}
In this work we proposed a method for data-driven (practical) stabilization of nonlinear systems  using nonparametric Chain Policies. The approach leverages a normalized  nearest-neighbor rule to assign, at each state, a finite-duration control signal, after which the process repeats. The method is grounded in the notion of Recurrent Lyapunov Functions (RLFs) as well as their control extension Control-RLFs, which enable certification of stability using standard norm function.

Our analysis establishes that:
\begin{enumerate}
    \item NPC Policies achieve practical exponential convergence to a $c$-neighborhood with sample complexity scaling as $O\!\left((3/\rho)^d\log(R/c)\right)$ in terms of the region radius $R$ and precision $c$.
    \item The framework supports incremental growth: new assignments can be added to expand the verified region while preserving previously established guarantees.
    \item Controller refinement is monotone: more data only improves convergence rates and enlarges certified region.
\end{enumerate}

These results position Chain Policies as a scalable, data-driven approach to certified stabilization, offering rigorous guarantees together with the ability to expand incrementally as new data becomes available.

\bibliographystyle{IEEEtran}
\bibliography{refs.bib}
\ifthenelse{\boolean{with-appendix}}{
\clearpage
\input{sections/appendix}
}{}
\end{document}